\documentclass[12pt]{article}
\usepackage[dvips]{graphicx}
\usepackage{latexsym}
\evensidemargin \oddsidemargin
\usepackage{amsmath}
\usepackage{epsfig}

\usepackage[T1]{fontenc} 
\usepackage[force]{feynmp-auto}
\usepackage{longtable}
\usepackage{psfrag} 

\usepackage{color}
\definecolor{orange}{RGB}{255,127,0}
\def\Comment#1{}
\newcommand{\bean}{\begin{eqnarray*}}
\newcommand{\eean}{\end{eqnarray*}}

\newcommand{\gapproxeq}{\lower
.7ex\hbox{$\;\stackrel{\textstyle >}{\sim}\;$}}
\newcommand{\lapproxeq}{\lower
.7ex\hbox{$\;\stackrel{\textstyle <}{\sim}\;$}}

\newcommand{\ie}{ {\it i.e.}, }
\newcommand\lsim{\mathrel{\rlap{\lower4pt\hbox{\hskip1pt$\sim$}}
    \raise1pt\hbox{$<$}}}
\newcommand\gsim{\mathrel{\rlap{\lower4pt\hbox{\hskip1pt$\sim$}}
    \raise1pt\hbox{$>$}}}
\newcommand{\ba}{\begin{array}}
\newcommand{\ea}{\end{array}}
\newcommand{\nn}{\nonumber}

\newcommand{\be}{\begin{equation}}
\newcommand{\ee}{\end{equation}}
\newcommand{\bear}{\begin{eqnarray}}
\newcommand{\eear}{\end{eqnarray}}

\newcommand{\ket}{\,\rangle}
\newcommand{\bra}{\langle \,}

\newcommand{\cO}{{\cal O}}

\newcommand{\mL}{\mathcal{L}}

\newcommand{\mF}{\mathcal{F}}

\newcommand{\mM}{\mathcal{M}}

\newcommand{\mT}{\mathcal{T}}

\newcommand{\Frac}[2]{\frac{\displaystyle #1}{\displaystyle #2}}
\newcommand{\Int}{\displaystyle{\int}}
\begin{document}


\thispagestyle{empty}

\def\thefootnote{\fnsymbol{footnote}}

\begin{flushright}
IFT-UAM/CSIC-14-028 \\
FTUAM-14-13
\end{flushright}

\vspace{0.5cm}

\begin{center}

{\large\sc {\bf One-loop $\gamma\gamma\to W^+_L W^-_L$ and $\gamma\gamma\to  Z_L Z_L$ from the}}

\vspace{0.4cm}

{\large\sc {\bf  Electroweak Chiral Lagrangian with a light Higgs-like scalar  }}

\vspace{1cm}

{\sc
R.L.~Delgado$^{1}$%
\footnote{email: rdelgadol@ucm.es  }%
,   A.~Dobado$^{1}$%
\footnote{email: dobado@fis.ucm.es  }%
, M.J.~Herrero$^{2}$%
\footnote{email: maria.herrero@uam.es}%
, J.J.~Sanz-Cillero$^{2}$%
\footnote{email: juanj.sanz@uam.es }%
}

\vspace*{.7cm}

{\sl
$^1$Departamento de F\'isica Te\'orica I, UCM, \\
Universidad Complutense de Madrid,
  Avda. Complutense s/n,
28040 Madrid, Spain

\vspace*{0.1cm}

$^2$Departamento de F\'isica Te\'orica and Instituto de F\'isica Te\'orica,
IFT-UAM/CSIC\\
Universidad Aut\'onoma de Madrid,
   C/ Nicol\'as Cabrera 13-15,
Cantoblanco, 28049 Madrid, Spain

}

\end{center}

\vspace*{0.1cm}

\begin{abstract}
\noindent
In this work we study the $\gamma\gamma\to W^+_L W^-_L$ and $\gamma\gamma\to  Z_L Z_L$ scattering processes within the effective chiral Lagrangian approach, including a light Higgs-like scalar as a dynamical field together with the would-be-Goldstone bosons $w^\pm$ and $z$ associated to the electroweak symmetry breaking. This approach is inspired by the possibility that the Higgs-like boson be a composite particle behaving as another Goldstone boson, and assumes the existence of a mass gap between $m_h$, $m_W$, $m_Z$ and the potential new emergent resonances, setting an intermediate energy region (above $m_{h,W,Z}$ and below the resonance masses) where the use of these effective chiral Lagrangians are the most appropriate tools to compute the relevant observables. We analyse in detail the proper chiral counting rules for the present case of photon-photon scattering  and  provide
the computation of the one-loop $\gamma\gamma\to W^+_L W^-_L$ and $\gamma\gamma\to  Z_L Z_L$
scattering amplitudes within this Effective Chiral Lagrangian approach and the Equivalence Theorem, including a discussion on the involved renormalization procedure. We also propose here a joint analysis of our results for the two-photon scattering amplitudes together with other photonic processes and electroweak (EW) precision observables for a future comparison with data. This could help to disentangle the nature of the light Higgs-like particle.
\end{abstract}

\def\thefootnote{\arabic{footnote}}
\setcounter{page}{0}
\setcounter{footnote}{0}

\newpage





 \tableofcontents

\section{Introduction}
The present consensus in the High Energy Physics Community points towards the interpretation that the recently discovered scalar particle at the CERN-LHC~\cite{Aad:2012tfa,Chatrchyan:2012ufa} could very well be the Higgs particle of the Standard Model of Particle Physics (SM). The most recent measurements of this scalar Higgs mass by the ATLAS and CMS collaborations set $m_h^{\rm ATLAS}=125.5\pm 0.6 $GeV~\cite{Aad:2013wqa} and $m_h^{\rm CMS}=125.7\pm 0.4 $GeV~\cite{Chatrchyan:2013lba}, respectively. These experiments also show that the most probable $J^P$ quantum numbers for this discovered particle are $0^+$, and conclude that the measured Higgs couplings to the other SM particles are in agreement so far, although yet with moderate precision, with the values predicted in the SM. Also the Higgs-like particle width $\Gamma_h$ has been found to be
 $\Gamma_h <  17.4$ MeV which is about 4.2 times the SM value~\cite{Width} .
However, there is one crucial issue of this discovered Higgs-like boson yet to be answered. The present data are still compatible with either an elementary or a composite Higgs boson hypothesis, therefore any optimal strategy to disentangle the real nature of this new scalar particle is very welcome in the search for a complete understanding of the Higgs system and the Higgs Mechanism of the Electroweak Theory.

In the present paper we assume that the Higgs boson is a composite particle and propose as one of these optimal strategies where
to look for deviations respect to the SM predictions, the particular scattering processes where two photons scatter and produce
two longitudinal weak bosons, i.e, we propose to look
at the scattering amplitudes $\mM(\gamma \gamma \to Z_L Z_L)$ and $\mM(\gamma \gamma \to W_L^+ W_L^-)$.
One of the most singular features of these two processes is that they do not receive contributions from the Higgs particle within the SM at the tree level. And even more, the neutral channel, $\gamma \gamma \to Z_L Z_L$ indeed vanishes in the SM at the tree level. Therefore, they are very sensitive processes to potential deviations from new physics in the Higgs sector, and they are specially appropriate for the case of a composite Higgs particle, since the new induced interactions from its composite nature with the photon-photon initial state could test the compositeness hypothesis more efficiently than other scattering processes not involving photons in the external legs.

Assuming that the Higgs boson is a composite particle, and not assuming any specific underlying strongly interacting theory explaining its properties in terms of its constituents, still allows for two qualitatively different possibilities. Either the composite Higgs particle appears as a resonance or it appears as a pseudo Goldstone Boson. On one hand we know that the measured Higgs mass is not far from the weak boson masses and, besides, there seems to be no new particles nor resonances in the spectra showing up at the presently available energies at LHC. This apparent mass gap between the boson masses, $m_W$, $m_Z$, $m_h$ and the potential new particles/resonances masses
     leads
to the preference for the hypothesis of the Higgs composite particle being another would-be-Goldstone-Boson. Notice
that the assumed Electroweak Symmetry Breaking pattern, $SU(2)_L \times U(1)_Y \to U(1)_{\rm em}$ seems to work, leading to the correct explanation of the mass generation for  $m_W$ and $m_Z$ with the three corresponding would-be-Goldstone-bosons (WBGBs), $w^{\pm}$ and $z$, transmuted
   into the three needed longitudinal components, $W^{\pm}_{L}$ and $Z_L$. Here we are inspired by the appealing idea that the composite Higgs boson, $h$, together with the $w^{\pm}$ and $z$ bosons are the associated Goldstone Bosons (GBs) of a {\it larger} spontaneous global symmetry breaking pattern containing the Electroweak Symmetry Breaking pattern, once the gauge interactions are included.

There are several Models proposed in the literature for a composite Higgs with specific implementations for the relevant global symmetry breaking pattern, like the Composite Higgs Model
based on the coset   $SO(5)/SO(4)$, usually called
minimally composite Higgs model (MCHM)~\cite{MCHM},
dilaton models with spontaneous breaking of scale invariance~\cite{dilaton}, and others~\cite{modelsreview}.
We do not consider here any specific model for composite Higgs, but instead work in a model independent way with the most appropriate tool provided by Effective Field Theories (EFTs). Concretely, we use here the EFT that is based in a non-linear realization of the Electroweak Chiral Symmetry Breaking (EWCSB) pattern, $SU(2)_L \times SU(2)_R \to SU(2)_{L+R}$, that is built with the so-called Electroweak Chiral Lagrangian with a light Higgs-like scalar (ECLh). The scalar sector of this ECLh contains the three needed $w^{\pm}$ and $z$ bosons and the Higgs particle $h$ as dynamical fields, and it shares with the SM the previously mentioned EWCSB pattern, with $SU(2)_C=SU(2)_{L+R}$ being the so-called custodial symmetry, and with the Higgs field being a singlet under this symmetry. The choice of a non-linear realization for the ECLh, instead of a linear one, and the name Chiral are due to the obvious close analogy with the Chiral Lagrangian (CL) of low energy QCD
 which, in its simplest version, is based in  the well known Chiral Symmetry Breaking pattern, $SU(2)_R \times SU(2)_L \to SU(2)_V$. In this QCD case the dynamics of mesons, that are identified with the associated GBs of this breaking, is well described by the CL~\cite{Weinberg:1979} and the systematic methods
of
Chiral Perturbation Theory (ChPT)~\cite{chpt}.
The choice of a non-linear realization for the QCD mesons is a crucial ingredient
for the understanding of the low energy meson phenomenology, in particular, for scattering
processes involving the dynamics of multiple mesons. This motivates the choice of a non-linear
realization also for the ECLh and, consequently, most of the EFT techniques
that were learnt from ChPT are nowadays applied to this new effective Lagrangian for the Electroweak Theory.

The ECLh is, on the other hand, the natural extension of the old models
for the Strongly Interacting Electroweak Symmetry Breaking Sector (SIEWSBS),
first introduced by Appelquist and Longhitano~\cite{Appelquist:1980vg,Longhitano:1980iz},
and later studied by other authors (see, for instance, \cite{Chanowitz:1985hj} and \cite{Cheyette:1987jf}). These old models for the SIEWSBS together with the methods inherited from the CLs~\cite{Weinberg:1979} and ChPT~\cite{chpt} for the study of low energy pion dynamics lead to the building of the so-called Electroweak Chiral Lagrangians (ECLs)~\cite{Dobado:1989ax}. These ECLs
were used mainly for
the study of the scattering of longitudinal EW gauge bosons~\cite{Dobado:1989ax,Dobado:1989ue,Dobado:1989gr,Dobado:1990jy,Pelaez:1995,Dobado:99}
and also for the study of other interesting observables like the ones proposed here,
$\mM(\gamma \gamma \to W_L^+ W_L^-)$ and $\mM(\gamma \gamma \to Z_L Z_L)$~\cite{Morales:92},
the so-called oblique $S$ and $T$ parameters~\cite{Peskin:92} or some related EW precision observables like $\Delta r$, $\Delta\rho$~\cite{Dobado:1990zh} and others~\cite{Espriu:1991vm}. These ECLs did not include the Higgs particle as an explicit dynamical field but instead
it was considered as a potentially emergent resonance of the strongly interacting underlying system. In fact the Higgs particle was assumed
to appear at the ${\cal O}(1 \,\,{\rm TeV})$ scale and there were indeed explicit computations of the ECL parameters emerging from
the integration to one-loop level of such
a heavy Higgs within the SM context~\cite{Morales:94,Morales:95}. After the discovery of a relatively light Higgs these ECLs are obviously not considered anymore, however, yet the generic EFT methods
that were developed for the ECLs, basically following a similar path as in ChPT, are yet applicable to the ECLh case.

At present  the complete list of  terms contributing to this ECLh at the tree level is well known,
including all the operators with bosonic and fermionic fields up to chiral dimension $d=4$~\cite{Alonso:2012,Brivio:2013},
and there are several works that use the ECLh for phenomenological purposes, like those devoted to the study of deviations
in the Higgs-like particle couplings to fermions and EW bosons at the tree level, and the possibility to disentangle
these deviations at the LHC~\cite{LHC-fits}.
On the other hand there are a few works including also the most relevant one-loop
contributions from the ECLh, like the ones studying
the scattering of longitudinal EW gauge bosons~\cite{Espriu:2013B,Dobado:2013},
the oblique $S$ and $T$ parameters~\cite{Pich:2013}
and others dealing with the renormalization program
of the effective action which, so far, has only been studied for the linear realization case~\cite{linear-EFT-Manohar}.

In this paper we present a one-loop computation of the $\mM(\gamma \gamma \to Z_L Z_L)$ and
$\mM(\gamma \gamma \to W_L^+ W_L^-)$ scattering amplitudes with the ECLh, valid up to next to leading order in the chiral expansion, and take a special attention to the role played by the hypothetical composite Higgs boson in these physical processes, as well as in the one-loop renormalization program with the ECLh that we describe also in detail here. We postpone the study of the potential experimental signatures at colliders, like LHC and future ILC, that are implied by our computation, for a future work.

For the explicit computation here with the ECLh we make the following assumptions and approximations:
\begin{itemize}
\item[1.-] The ECLh is $SU(2)_L \times U(1)_Y$ gauge invariant and provides a good description of scattering processes, whenever their relevant energies, $\sqrt{s}$, are sufficiently low, say,
$\sqrt{s} \ll \Lambda_{\rm ECLh}$,
where $\Lambda_{\rm ECLh}={\rm min} \left( 4\pi v, M \right)$. Here, $v=246 \,\,{\rm GeV}$, $4\pi v \sim 3\,\,{\rm TeV}$ is the typical scale introduced by the chiral loops, as in any other Chiral Lagrangian, and $M$ refers to the mass of any potentially emergent resonance. In the present paper we focus on the bosonic part of the ECLh and leave the fermionic contributions for another work. We also assume CP invariance in our selection of the relevant terms of the ECLh.
\item[2.-] We work in the Landau gauge that simplifies the one-loop computation since it implies massless WBGBs, i.e., for the present computation we take $m_{w^{\pm}}=m_z=0$.
\item[3.-] We use the Equivalence Theorem (ET)~\cite{equivalence-theorem}
that has been proven to work also
within the context of a SIEWSBS~\cite{ET+ECL}.
For the present computation it means that, for energies well above the EW gauge boson masses, $ m_W, m_Z \ll \sqrt{s}$, the following approximations can be done:
	 \begin{eqnarray}
	 \mM(\gamma \gamma \to W_L^+ W_L^-)& \simeq &-\mM(\gamma \gamma \to w^+ w^-)  \\
	 \mM(\gamma \gamma \to Z_L Z_L)& \simeq &-\mM(\gamma \gamma \to z z)
	 \end{eqnarray}
The use of this theorem will allow us to extract
the leading contributions to our observables  in terms
of diagrams with only $w^{\pm}$, $z$ and $h$ in the internal lines. The diagrams with
internal $\gamma, W^\pm,Z$  lines
would enter at higher orders in $g$ and $g'$ and these would lead to subleading contributions under the assumption of  small $g,g'$, or more precisely for
$gv,\, g^{'} v \ll \sqrt{s}$ with $v$  kept fixed, which is precisely the energy range set by the ET,
$m_W, m_Z \ll \sqrt{s}$. Besides, due to the close values of $m_h$ with $m_W$ and $m_Z$, the  two previous assumptions together lead to the following window of applicability in energies for our computation:
       \begin{equation}
       m_h, m_W, m_Z \ll \sqrt{s} \ll  \Lambda_{\rm ECLh}
       \label{energyrange}
        \end{equation}

       \item[4.-] Finally, we also assume custodial symmetry invariance in the scalar sector of the ECLh as in the SM. It means that
       the custodial symmetry breaking terms included here for the bosonic sector are exclusively induced by the gauging of the hypercharge group, $U(1)_Y$, hence their corresponding contributions to the observables will be driven by the 'small' coupling $g'$. Equivalently, by setting to zero the hypercharge gauge coupling in the ECLh considered here one recovers the full symmetry, $SU(2)_L\times SU(2)_R$.
 \end{itemize}
The paper is organized as follows. In section~\ref{sec.Lagrangian}
we introduce the ECLh
and select the relevant terms for the present computation
of the one-loop $\mM(\gamma \gamma \to w^+ w^-)$ and $\mM(\gamma \gamma \to z z)$
scattering amplitudes up to ${\cal O}(e^2p^2)$. We also include a detailed description on how we
implement the chiral dimension counting in the ECLh.
The relevant chiral parameters for the present computation will also be specified in this section.
Section.~\ref{sec.counting}   contains an illustrative and generic description of the various contributions entering in the present computation, analyzing separately
the Leading Order (LO) tree level terms, i.e. of ${\cal O}(e^2)$, the Next to Leading Order tree level
terms, i.e. of ${\cal O}(e^2p^2)$, and the one-loop contributions generated from the LO terms, which are of this same order, i.e. of ${\cal O}(e^2p^2)$.
The renormalization of the ECLh parameters is also discussed in this section.
Section~\ref{sec.coset} introduces the two different parametrizations of the coset
that we have used in this computation, as a check of the parametrization independence of our final result. We also present the final effective Lagrangian in terms of the relevant fields,
the photon field and the $w^{\pm},z,h$ fields, and derive the corresponding  Feynman Rules
(collected in Appendix~\ref{app.Feynman}) for the two chosen parametrizations.
Section~\ref{sec.results} contains the  analytical results of our computation of the one-loop $\gamma \gamma \to w^aw^b$ scattering amplitudes .
In section~\ref{sec.discusion} we discuss the results and compare them
with the predictions of other interesting observables that have been selected
here because they involve the same ECLh parameters and therefore there are correlations among them. We also propose here the use of global analysis for all these observables as a promising method to extract the values of the
ECLh parameters from data. Section~\ref{sec.conclusions}
is finally devoted to the conclusions.
Technical details as the Feynman rules, the individual contributions from the various loop diagrams, the detailed predictions for the related observables and the specific results for the $\gamma \gamma$ scattering processes in the particular model MCHM have been
relegated to the appendices.

\section{The Electroweak Chiral Lagrangian with a light Higgs}
\label{sec.Lagrangian}

The ECLh is a gauged non-linear effective Lagrangian coupled to a singlet scalar particle that contains as dynamical fields the EW gauge bosons, $W^{\pm}$, $Z$ and $\gamma$, the corresponding would-be GBs, $w^{\pm}$, $z$, and the Higgs-like scalar boson, $h$.  The WBGBs, $w^{\pm}$, $z$, are described by a matrix field $U$ that takes values in the $SU(2)_L \times SU(2)_R/SU(2)_{L+R}$ coset,   and transforms as $U \to L U R^\dagger$ under the action of the global group  $SU(2)_L \times SU(2)_R$ that defines the EW Chiral symmetry. The subgroup $SU(2)_{L+R}=SU(2)_C$ defines the custodial symmetry group. We will assume here that, as it happens in the SM,  the scalar sector of the ECLh preserves this custodial symmetry, except for the explicit breaking due to the gauging of the $U(1)_Y$ symmetry. Two particular parametrizations of this unitary matrix $U$ in terms of the dimensionless $w^{\pm}/v$ and $z/v$ fields, with $v=246\,\,{\rm GeV}$ will be presented in section 4, where it will be also commented on the advantages and disadvantages of each of these parametrizations,  both leading to the same predictions for the physical observables. Regarding the Higgs field $h$ one has to take into account that it is a singlet under the EW Chiral symmetry $SU(2)_L \times SU(2)_R$ and, therefore, there are not particular restrictions on the implementation of this field into the ECLh from the EW Chiral symmetry requirements. The Higgs field is consequently introduced in the ECLh via multiplicative polynomial functions, ${\cal F}(h)$, and their derivatives. These polynomials are generically of the form,
$(1+2ah/v+b (h/v)^2+...)$, where $a,b..$ are general ECLh parameters that are added to the other ECLh parameters and that take particular values in specific models. The EW gauge fields are introduced as usual by means of the gauge principle that
ensures the $SU(2)_L\times U(1)_Y$ gauge invariance of the ECLh. Concretely, they are introduced by the covariant derivative of the $U$ field, $D_\mu U$, and by the $SU(2)_L$ and $U(1)_Y$ field strength tensors, $\hat{W}_{\mu \nu}$ and $\hat{B}_{\mu\nu}$ respectively. In summary, the basic functions for the building of the
$SU(2)_L\times U(Y)_L$ gauge invariant ECLh are the following:
\begin{eqnarray}
U(w^\pm,z) &=& 1 + i w^a \tau^a/v + \cO(w^2)\;\in SU(2)_L \times SU(2)_R/SU(2)_{L+R}, \label{Umatrix}\\
{\cal F}(h)&=& 1+2a\frac{h}{v}+b \left(\frac{h}{v} \right)^2+\dots ,\label{polynomial}\\
D_\mu U &=& \partial_\mu U + i\hat{W}_\mu U - i U\hat{B}_\mu, \label{eq.cov-deriv} \\
\hat{W}_{\mu\nu} &=& \partial_\mu \hat{W}_\nu - \partial_\nu \hat{W}_\mu
 + i  [\hat{W}_\mu,\hat{W}_\nu ],\;\hat{B}_{\mu\nu} = \partial_\mu \hat{B}_\nu -\partial_\nu \hat{B}_\mu ,\label{fieldstrength}\\
\hat{W}_\mu &=& g \vec{W}_\mu \vec{\tau}/2 ,\;\hat{B}_\mu = g'\, B_\mu \tau^3/2 ,
\label{EWfields}\\
V_\mu &=& (D_\mu U) U^\dagger ,\; \mT = U\tau^3 U^\dagger ,\label{VmuandT}
\end{eqnarray}
where we have also included the usual
$V_\mu$ and  $\mT$ chiral fields.

According to the usual counting rules of Chiral Lagrangians, the $SU(2)_L \times U(1)_Y$ invariant terms in the ECLh are organized by means of their 'chiral dimension', meaning that a term ${\cal L}_d$ with
'chiral dimension' $d$ will contribute to $O(p^d)$ in the corresponding power momentum expansion. For the present computation of one-loop
$\mM(\gamma \gamma \to W_L^+ W_L^-)$ and $\mM(\gamma \gamma \to Z_L Z_L)$ scattering amplitudes
by means of the Equivalence Theorem there are just a few terms in the
ECLh that are involved in the corresponding one-loop
$\mM(\gamma \gamma \to w^+ w^-)$ and $\mM(\gamma \gamma \to zz)$
scattering amplitudes. Thus, we focus here mainly on this subset of ECLh terms that we present
in this section and classify according to its chiral dimension.

The chiral dimension of each term in the ECLh can be
found out by following the scaling with $p$ of the various contributing basic
functions. First, as it is usual in Effective Chiral Lagrangians, the derivatives and the masses of the dynamical particles are considered as soft scales of the EFT and are consequently of the same order in the chiral counting, i.e. of ${\cal O}(p)$. The gauge boson masses, $m_W$ and $m_Z$ are    examples of these soft masses in the case of the ECLh. These are generated from the covariant derivative in
Eq.~(\ref{eq.cov-deriv}) once the $U$ field is expanded in terms of the $w^a$ fields as:
\bear
D_\mu U &=& \Frac{i \partial_\mu \vec{w}\, \vec{\tau}  }{v}
\quad +\quad i\,  \Frac{g v}{2}  \, \Frac{\vec{W}_\mu\, \vec{\tau}}{v}
\quad -\quad i\,  \Frac{g' v}{2}  \, \Frac{B_\mu\,  \tau^3}{v}
\quad +\quad ...
\eear
where the dots represent terms with higher powers of $(w^a/v)$ and whose precise form will depend on the particular parametrization of $U$.  Once the gauge fields are rotated to the physical basis then they get the usual gauge boson squared mass values at lowest order: $m_W^2=g^2 v^2/4$ and $m_Z^2 =(g^2+g^{'2}) v^2 /4$. Furthermore, in order to have a power counting consistent with
the loop expansion one needs
all the terms in the covariant derivative above to be of the same order. Thus, the
   proper   assignment is
 $\partial_\mu\, ,(g v)\, , (g'v) \sim \cO(p)$ or, equivalently, $\partial_\mu\, ,m_W\, , m_Z \sim \cO(p) $. In addition, due to the close values of the EW gauge boson masses with the experimental Higgs boson mass,   we
   will
 also consider in this work the Higgs-like boson mass $m_h$ as another soft mass in the ECLh with, a
similar chiral counting as $m_W$ and $m_Z$.  That implies, $m_h \sim
\cO(p)$, or equivalently $(\lambda v^2) \sim \cO(p^2)$, with $\lambda$ being the SM Higgs self-coupling. One can similarly conclude on the scaling of all the other building blocks of the
ECLh that we summarize and collect in the following:
\bear
\partial_\mu,\; m_W,\; m_Z,\; m_h &\sim & \cO(p), \label{eq.mW-mZ-p-scaling} \\
D_\mu U,\; V_\mu,\; g'v,\; \mT,\; \hat{W}_{\mu},\; \hat{B}_{\mu} &\sim & \cO(p), \label{eq.DU-Vmu-scaling}\\
\hat{W}_{\mu\nu},\;\hat{B}_{\mu\nu} &\sim & \cO(p^2).
\label{eq.Wmunu-Bmunu-scaling}
\eear
Notice that to get the correct chiral counting of quantities involving the couplings $g$ and/or $g'$ it is convenient to rewrite them in terms of $(g v)$ and/or $(g' v)$ and dimensionless fields, correspondingly. For instance,  $\hat{B}_{\mu\nu}=(g'v)\partial_\mu (B_\nu/v)\tau^3/2-(g'v)\partial_\nu (B_\mu/v)\tau^3/2 \sim \cO(p^2)$. Similarly, one can check other examples like $(1/g')^2\hat{B}_{\mu\nu} \hat{B}^{\mu\nu}\sim \cO(p^2)$ etc.

With these building blocks one then construct the ECLh up to a given order in the chiral expansion. We require this Lagrangian to be CP invariant, Lorentz invariant and $SU(2)_L \times U(1)_Y$ gauge invariant. For the present work we include terms with chiral dimension up to  $\cO(p^4)$, therefore, the ECLh can be generically written as:
\be
\mL_{\rm ECLh} = \mL_2 + \mL_4 +\mL_{\rm GF} +\mL_{\rm FP}\, ,
\ee
where  $\mL_2$ refers to the terms with chiral dimension 2, i.e $\cO(p^2)$,  $\mL_4$ refers to the terms with chiral dimension 4, i.e $\cO(p^4)$, and
$\mL_{\rm GF}$ and $\mL_{\rm FP}$ are the gauge-fixing (GF) and the corresponding non-abelian Fadeev-Popov (FP)
terms that have been explicitly separated as they are particular terms added to
EW gauge theory  to fix the gauge freedom in the path integral.
As we said in the introduction, the Landau gauge is assumed all along this work,
 which is the most convenient one for the present computation since the WBGBs $w^\pm$ and $z$ are massless in this gauge. The convenience of the Landau gauge choice in the context of the
gauged non-linear sigma model was emphasized long ago in \cite{Appelquist:1980vg}, since in this gauge there are no direct couplings of the GB to the ghosts.

We focus next in the relevant terms for $\gamma \gamma \to w^+ w^-$ and  $\gamma \gamma \to z z$ scattering processes. First, the relevant terms in the leading order (LO) Lagrangian --of $\cO(p^2)$-- are given by
\bear
\mL_2 &=&    -\Frac{1}{2 g^2} {\rm Tr}(\hat{W}_{\mu\nu}\hat{W}^{\mu\nu}) -\Frac{1}{2 g^{'2}} {\rm Tr} (\hat{B}_{\mu\nu} \hat{B}^{\mu\nu})\nn\\
&& +\Frac{v^2}{4}\left[%
  1 + 2a \Frac{h}{v} + b \Frac{h^2}{v^2}\right] {\rm Tr} (D^\mu U^\dagger D_\mu U )
 + \Frac{1}{2} \partial^\mu h \, \partial_\mu h + \dots\, ,
\label{eq.L2}
\eear
where the dots stand for $\cO(p^2)$ operators with three
or more  Higgs fields which do not enter into this computation. In particular, notice that the Higgs self-interaction terms do not enter here for this same reason. Besides, the Higgs mass term is not included either, because it would lead to subleading contributions to the observables of interest here when compared to the set of contributions that we are considering which will dominate in the assumed energy range given in Eq.~(\ref{energyrange}).

Second, the complete next to leading order (NLO) Lagrangian --of $\cO(p^4)$--  contain many terms. In particular it includes the complete set of CP-even, Lorentz and gauge invariant operators that were first collected by
Longhitano in ref.~\cite{Longhitano:1980iz},
and also listed in other works
like~\cite{Pelaez:1995,Dobado:99,Morales:92,Dobado:1990zh,Morales:94,Morales:95}
and references therein.    
The total list in ref.~\cite{Longhitano:1980iz} includes 14 operators of chiral dimension 4, which are reduced to just 5 if one rejects operators that are not custodial symmetry invariant, even in the case of switching off the $g'$ gauge couplings, and also by the use of the equations of motion.
This reduced list of 5 terms is given by the explicit terms below:
\bear
{\cal L}_{\rm Longhitano} &=& %
  a_1 {\rm Tr}( U \hat{B}_{\mu\nu} U^\dagger \hat{W}^{\mu\nu})
  + i a_2  {\rm Tr} ( U \hat{B}_{\mu\nu} U^\dagger [V^\mu, V^\nu ])
  - i a_3 {\rm Tr} (\hat{W}_{\mu\nu}[V^\mu, V^\nu]) \nn \\
&&
  + a_4 \left[{\rm Tr}(V_\mu V_\nu) \right] \left[{\rm Tr}(V^\mu V^\nu)\right]
  + a_5 \left[{\rm Tr}(V_\mu V^\mu) \right] \left[{\rm Tr}(V_\nu V^\nu)\right]+\dots\, ,
\eear
where our notation for the chiral parameters $a_i$ is as in refs.~\cite{Morales:94,Morales:95} and
   they are
related to Longhitano's $\alpha_i$ chiral parameters by: $a_1=(g/g') \alpha_1$, $a_2=(g/g') \alpha_2$,
$a_3= -\alpha_3$, $a_4= \alpha_4$, $a_5= \alpha_5$.
They are also related with the usual notation for the chiral parameters $L_i$  from
the QCD Chiral Lagrangian~\cite{chpt} restricted to two light flavours~\footnote{
Notice that while this correspondence between the ECL parameters $a_i$
and the $SU(3)$ CL parameters $L_j$  from QCD
is valid at the tree-level, this is no longer true at the loop level where the renormalization and running of the effective Lagrangian parameters
depend on the symmetry group.
However, we can still keep the analogy between the EW and QCD effective Lagrangians at the loop level if
we relate the ECL parameters with those in the original CL for low energy QCD in the case of $SU(2)$~\cite{chpt} leading to:
$\ell_1=4 a_5$, $\ell_2= 4 a_4$, $\ell_5= a_1$, $\ell_6= 2(a_2 -a_3)$.}:
$L_1=a_5$, $L_2=a_4$, $L_9=a_3-a_2$, $L_{10}=a_1$.

The parameters $a_4$ and $a_5$, on the other hand,  are of great relevance since they participate with a leading role in the scattering of longitudinal EW gauge bosons like, for instance,
$W_L^+W_L^- \to W_L^+ W_L^-$ and  $W_L^+W_L^- \to Z_L Z_L$. These have been studied within the ECLh context by several authors~\cite{Espriu:2013B,Dobado:2013} and they are of much interest due to the implications for LHC which presumably will explore these scattering amplitudes in the future run. However the implications of  $a_{1,2,3}$ within the context of the ECLh have not been studied yet, and this is one of our goals here. As we will see next, these $a_{1,2,3}$ are the relevant chiral parameters entering in $\gamma \gamma \to w^a w^b$ scattering.
Notice also that the first three operators involving $a_{1,2,3}$ have been written
in such a way that if the  $U(1)_Y$ gauge group
were promoted to a wider symmetry $SU(2)_R$ with the help of spurionic fields for the
two missing gauge bosons, these operators
would be $SU(2)_L\times SU(2)_R$ invariant.

Finally, in the building of the $O(p^4)$ terms of the ECLh, one has to add extra terms involving the Higgs field, which include adding polynomial factors in front of the previous operators of the generic type $(1+k_i (h/v)+g_i (h/v)^2+..)$ and also in front of the other $O(p^4)$ terms like $\hat{W}_{\mu\nu} \hat{W}^{\mu\nu}$ and   $\hat{B}_{\mu\nu} \hat{B}^{\mu\nu}$. In summary, by selecting the subset of    $O(p^4)$ terms that are relevant for the scattering of interest here, i.e. for
$\gamma \gamma \to w^+w^-$ and $\gamma \gamma \to zz $, we find the following
short list of contributing terms:
\bear
\mL_4 &=&
  a_1 {\rm Tr}(U \hat{B}_{\mu\nu} U^\dagger \hat{W}^{\mu\nu})
  + i a_2 {\rm Tr} (U \hat{B}_{\mu\nu} U^\dagger [V^\mu, V^\nu ])
  - i a_3  {\rm Tr} (\hat{W}_{\mu\nu}[V^\mu, V^\nu]) \nn\\
&&
  - c_{W} \Frac{h}{v} {\rm Tr}(\hat{W}_{\mu\nu} \hat{W}^{\mu\nu})
  - c_B \Frac{h}{v}\, {\rm Tr} (\hat{B}_{\mu\nu} \hat{B}^{\mu\nu} ) + \dots
\label{eq.L4}
\eear
where  we have used the same conventions for the definition of the $c_W$ and $c_B$ parameters
as in Refs.~\cite{Alonso:2012,Brivio:2013}.
Notice that the last two terms in the previous equation, once the gauge fields are rotated to the physical basis,
lead to one of the relevant operators here involving the photon
field strength, $A_{\mu \nu}=\partial_\mu A_\nu- \partial_\nu A_\mu$, which is:
\be
-c_W\Frac{h}{v}{\rm Tr}(\hat{W}_{\mu\nu} \hat{W}^{\mu\nu})
-c_B\Frac{h}{v}{\rm Tr}(\hat{B}_{\mu\nu} \hat{B}^{\mu\nu})
=
-\Frac{c_{\gamma}}{2}\Frac{h}{v}e^2 A_{\mu\nu} A^{\mu\nu}
+\dots
\ee

Finally to finish this section, we find illustrative to compare the different settings for the ECLh parameters in some specific
and most popular models, like the Higgsless ECL models, the SM itself,  the $SO(5)/SO(4)$ MCHM,
   dilaton   models and others. If we make the comparison at the LO Lagrangian level (i.e at ${\cal L}_2$ level), the values of the ECLh parameters in these models are, correspondingly:
\bear
& a^2=b=0
&\qquad
\mbox{Higgsless ECL~\cite{Appelquist:1980vg,Longhitano:1980iz},}\nn\\
& a^2=b=1
&\qquad
\mbox{SM,}
\nn\\
&
a^2=1-\Frac{v^2}{f^2},\,
b=1-\Frac{2v^2}{f^2}
&\qquad
\mbox{SO(5)/SO(4)   MCHM~\cite{MCHM},}
\nn\\
&
a^2=b=\Frac{v^2}{\hat{f}^{2}},
&\qquad
\mbox{Dilaton~\cite{dilaton}.}
\label{models}
\eear

If we make the comparison at the NLO Lagrangian level, then one has to set in addition the values of the  ECLh parameters in ${\cal L}_4$. Thus, for instance, in comparing with the Higgsless ECL models one sets, in addition to the previous values above, $c_\gamma$ and all the involved parameters in the polynomial functions for the Higgs field to zero. The rest of $a_i$ parameters are present in the ECL as in the ECLh case. In the comparison with the SM, for consistency, one has to set obviously all parameters in ${\cal L}_4$ to zero, $a_i=0$, $c_\gamma=0$, etc.

\section{Electroweak Chiral Loops and renormalized ECLh parameters}
\label{sec.counting}

In this section we describe the systematic procedure for our computation of the one-loop
$\mM(\gamma \gamma \to w^+w^-)$ and $\mM(\gamma \gamma \to z z)$ scattering amplitudes, starting with the relevant terms of the
ECLh  that have been fixed in the previous section.

First, as in any Chiral Effective Theory, one has to fix the order in the Chiral expansion up to which the amplitude is to be computed. Here we set the computation up to ${\cal O}(p^4)$, meaning that the amplitude will have two type of contributions: the Leading Order contributions (LO), i.e, ${\cal O}(p^2)$ and the Next to leading order contributions (NLO), i.e, ${\cal O}(p^4)$. Next, and following the
standard counting rules of
 Chiral Lagrangians~\cite{Weinberg:1979,chpt,chpt+photons},
one has to consider the contributions from chiral loops here called, for obvious reasons, electroweak chiral loops, that are produced by the LO Lagrangian.  In the present context of ECLh these EW chiral loops will include the Higgs boson particle in the loops, in addition to the usual dynamical fields of the Chiral Lagrangians. These EW chiral loops do also contribute to order ${\cal O}(p^4)$ and have to be taken into account together with the fixing of a well defined renormalization prescription to deal with the divergences that
  are generated by the EW chiral loops, usually computed in dimensional regularization which will also be considered in this work. This implies setting a well defined procedure for the renormalization of the ECLh parameters. Notice that dimensional regularization is particulary appropriate for dealing with CL since it is chiral invariant so that no extra terms must be added to the action to restore chiral invariance as would be the case if one uses a cutoff or other non-invariant regularization methods (see for example \cite{mybook} and references therein).

In order to clarify the various steps to follow in our case, let us first shortly review how do the Weinberg's chiral power counting rules~\cite{Weinberg:1979} apply to the present case of ECLh. With this purpose in mind, let us first write the
term with chiral dimension $d$ in the ECLh, ${\cal L}_d$, in the generic form:
\be
\mL_d = \sum_k f_k^{(d)} p^{d}\left(\Frac{\chi}{v}\right)^k,\label{eq.Ld-symbolic}
\ee
where, $\chi$ refers to any of the bosonic fields in the ECLh, $h$, $w^a$, $W^a_\mu$, $B_\mu$; $p$ refers to either a derivative $\partial$ acting on the corresponding bosonic field or a soft mass,
$m_{W,Z,h}$; and $f_k^{(d)}$ are the corresponding
   coefficients       
in front of the terms with $k$ powers of the dimensionless field $(\chi/v)$. For instance, for the lowest dimension terms, $\mL_2$ and $\mL_4$, these go respectively as: $f_k^{(2)} \sim v^2$ and
$f_k^{(4)} \sim a_i$, where by $a_i$ we mean here generically all the dimensionless parameters in $\mL_4$ like
$a_1$, $a_2$, $a_3$ and $c_\gamma$. The coefficients of higher chiral dimension terms $f_k^{(d)}$ will be correspondingly of lower energy (canonical) dimension $[E]^{4-d}$. Now, according to the usual Weinberg's chiral counting rules, the contribution to a given scattering amplitude from each Feynman diagram containing
$L$ loops generated with the LO $\mL_2$, $E$ external legs and $N_d$ interaction vertices from $\mL_d$, scales in powers of $p$
as follows:
\be
\mM\sim\left(\Frac{p^2}{v^{E-2}}\right)\left(\Frac{p^2}{16\pi^2 v^2}\right)^L\prod_d \left(\Frac{f_k^{(d)} p^{(d-2)}}{v^2}\right)^{N_d},
\label{eq.chiral-counting}
\ee
where we see explicitly the typical suppression factor of the chiral loops given by $p^2/(16\pi^2v^2)$
for each loop.   For further details on the power counting in EW Chiral Lagrangians,   see
Refs.~\cite{ECLh-Cata,EW-chiral-counting}.
Notice also that, for $d=2$, the last factor scales as $p^0$, meaning that the same result for the total scaling of
$\mM$     
is obtained for any number $N_2$ of vertices from ${\cal L}_2$ entering in the Feynman diagram.
Then, applying this chiral counting to our  $\gamma\gamma\to w^a w^b$ scattering processes (with four external legs), we find immediately that the contribution from a generic Feynman diagram containing $L$ loops generated with $\mL_2$, and $N_4$ interaction vertices from $\mL_4$ (and for any $N_2$) scales in powers of $p$ as follows:
\be
\mM\sim\left(\Frac{p^2}{v^2}\right)\left(\Frac{p^2}{16\pi^2 v^2}\right)^L
\left(\Frac{a_i p^2}{v^2}\right)^{N_4}\sim(e^2)\left(\Frac{p^2}{16\pi^2 v^2}\right)^L
\left(\Frac{a_i p^2}{v^2}\right)^{N_4},
\label{eq.chiral-counting4}
\ee
where, as already stated above, the $a_i$ here refer generically to all the EW chiral parameters of $\mL_4$, i.e., $a_1$, $a_2$, $a_3$ and $c_\gamma$ in the present work. Furthermore, for the particular present case with two photons in the initial state, as it will be shown later in the explicit computation, the first factor $(p^2/v^2)$ indeed appears here as $e^2$ with $e$ being the electromagnetic coupling, therefore, we have also rewritten in Eq.(\ref{eq.chiral-counting4}) this generic expression in terms of the $(e^2)$ pre-factor.

Summarizing, we can now use the previous expression to the right in Eq.~\eqref{eq.chiral-counting4} to conclude on the various contributions entering into our computation of the one-loop $\mM(\gamma\gamma\to w^a w^b)$ amplitudes, and also set the various steps to follow. Generically, we will split the amplitudes into two parts:
\begin{equation}
\mM=\mM_{\rm LO}+\mM_{\rm NLO},
\end{equation}
where:
\begin{enumerate}
\item The LO contributions are given by all the tree-level diagrams ($L=0$)
with vertices from just the $\mL_2$ Lagrangian, i.e with $N_4=0$:
\be
\mM_{\rm LO}=
\mM^{\rm tree}_{\cO(e^2)}    
\sim e^2.
\ee

\item The NLO contributions come from two types of graphs:
\be
\mM_{\rm NLO} =
\mM^{\rm 1-loop}_{\cO(e^2 p^2)}   
+
\mM^{\rm tree}_{\cO(e^2 p^2)}    
\label{NLO}
\ee

\begin{itemize}

\item One-loop diagrams ($L=1$)
with only vertices  from $\mL_2$, i.e with $N_4=0$. These contribute as,
\be
\mM^{\rm 1-loop}_{\cO(e^2 p^2)}      
\,\,\, \sim\,\,\, e^2\,\,\, \left(\Frac{p^2}{16\pi^2 v^2}\right) \, .
\ee
It is known that these loops may generate  UV divergences, therefore,
requiring the presence of local counter-terms to fulfill the renormalization.

\item Tree-level diagrams ($L=0$) with only one vertex  from  the $\mL_4$ Lagrangian, i.e. with $N_4=1$, and the remaining vertices from  the $\mL_2$ Lagrangian. These contribute as,
\be
\mM^{\rm tree}_{\cO(e^2 p^2)      
}\,\,\, \sim\,\,\,  e^2\,\,\, \left(\Frac{a_i\, p^2}{v^2}\right) \, .
\ee

\end{itemize}
\item The final step is to define a specific renormalization prescription to cancel out the one-loop UV divergences that appear in $\mM^{\rm 1-loop}_{\cO(e^2 p^2)}$. This is performed in a systematic way by the renormalization of the EW chiral parameters in ${\cal L}_4$. In contrast, the parameters in ${\cal L}_2$ do not get renormalized.
This renormalization program is completely analogous to the one in
Chiral Perturbation Theory~\cite{Weinberg:1979,chpt}. In our present case this will imply, in principle, the renormalization of the four  EW chiral parameters entering here, namely, $a_1$, $a_2$, $a_3$ and $c_\gamma$:
\be
a_1, a_2, a_3, c_\gamma \to a_1^r, a_2^r, a_3^r, c_\gamma^r
\ee
by the proper counterterms $\delta a_i$, with
\begin{equation}
a_i^r=a_i+ \delta a_i,
\end{equation}
that will remove the UV divergences appearing, as usual in dimensional regularization, as functions of
    $1/\hat{\epsilon}$
defined as,
\be
\frac{1}{\hat{\epsilon}}=\mu^{-2\epsilon}
\left(\frac{1}{\epsilon} -\gamma_E +\ln{4\pi}\right),
\label{epsilon}
\ee
 with  $D=4-2\epsilon$.
These $\delta a_i$ counterterms in turn, will lead to the corresponding running of the
effective EW parameters, i.e. the dependence with the renormalization scale $\mu$ of these parameters, $a_i^r(\mu)$.

One of the big surprises in this work, that we anticipate here, is that after the
very intriguing computation of all the very many loops and contributing terms to the
$\gamma \gamma$ scattering amplitudes, it turns out that the final result for the
$\mM^{\rm 1-loop}_{\cO(e^2 p^2)}$
terms are indeed finite!!. This means that the particular  combination   
of the EW chiral parameters
$a_1, a_2, a_3, c_\gamma$ entering here does not need renormalization, and therefore this will give us the result for the physical one-loop amplitude in terms of a renormalization group invariant combination of these $a_1, a_2, a_3, c_\gamma$ parameters. This is a very interesting result, but all this will be presented and discussed in full detail later.
\end{enumerate}

\section{Coset parametrizations and relevant Feynman rules}
\label{sec.coset}

In order to perform the computations  we are dealing with in this work we have to chose some parametrization of the coset. As it was described in the previous section our effective low-energy theory is a gauged non-linear sigma model based in the coset $SU(2)_L \times SU(2)_R/ SU(2)_{L+R}$ coupled to the light scalar $h$. The gauge group is $SU(2)_L\times U(1)_Y$ and $SU(2)_{L+R}$ is the custodial symmetry group.  Therefore the coset is just the space $SU(2)$
which  is   isomorphic to the three dimensional sphere $S^3$.
In order to have an explicit effective Lagrangian we need to introduce some particular coordinates on this space that will play the role of the WBGB fields. These three  fields must be independent and properly normalized but otherwise they are arbitrary since there is a well known theorem about non linear sigma models guaranteeing  that the $S$ matrix elements (on-shell amplitudes) are independent of the particular  coordinates chosen
(see for example \cite{mybook} and references therein). For the case considered here, one of the most popular elections is the exponential representation given by:
\bear
U(x) &=& \exp i\frac{\tilde{\pi}}{v}\, ,
\eear
where $\tilde{ \pi}= \tau^a \pi^a(x)$ and $\tau^a$ $(a=1,2,3)$ are the Pauli matrices. However, as we will see later, this is not at all the most efficient coset parametrization in this case. These coordinates  are inspired in ChPT where one usually deals with a $SU(3)$ coset. However here we are using just a $SU(2)$ coset which is isomorphic to $S^3$. Then it is possible to introduce the much simpler coordinates:
\bear
U(x) &=& \sqrt{1-\frac{\omega^2}{v^2}}+ i \frac{\tilde\omega}{v} \, ,
\eear
where again $\tilde{ \omega}= \tau^a \omega^a(x)$ and $\omega^2=\sum_a (\omega^{a})^2=\tilde \omega ^2$. We will call these coordinates spherical to distinguish them from the exponential ones. As we will see next, the Lagrangian, the  Feynman rules and even the Feynman diagrams are simpler
(meaning lesser in number)  in the spherical representations but the final results for  the amplitudes are parametrization independent, as expected. We have checked explicitly this fact by making the computations using both representations independently.

It is not very difficult to find the transformation equations passing from one set of coordinates to the other. To do that we first realize that the
exponential representation can also be written
as:
\be
U(x) = \cos \frac{\pi}{v}+ i \frac{\tilde \pi}{\pi} \sin \frac{\pi}{v},
\ee
where $\pi =\sqrt{\pi^2}$ with $\pi^2=\sum_a (\pi^{a})^2$. Therefore,  we have:
\be
\omega^a= \pi^a \frac{v}{\pi}\sin  \frac{\pi}{v},
\ee
which implies $\omega^2= v^2 \sin^2(\pi/v)$. By expanding the result above we can get the series:
\be
\omega^a= \pi^a \left[1-\frac{1}{6}\bigg(\frac{\pi}{v}\bigg)^2
+\frac{1}{120}\bigg(\frac{\pi}{v}\bigg)^4-\frac{1}{5040}\bigg(\frac{\pi}{v}\bigg)^6+\dots
\right].
\ee
Also it is easy to show the useful relation
\be
\frac{\omega^a}{\omega}=\frac{\pi^a}{\pi}.
\ee
{}From these fields it is useful to introduce the fields $\omega^{\pm}=(\omega^1   \mp  i \omega^2)/\sqrt {2}$, $ \omega^0 = \omega^3$
which implies $\omega^2=2\omega^+\omega^-+\omega^0\omega^0$
 and similar definitions for $\pi^{\pm}$ and $\pi^0$.
These two sets of fields represent equally the WBGB   $w^{\pm}$ and $z$ responsible for the  $W^{\pm}$ and $Z$ masses respectively.

Now we can write the leading order Lagrangian found in the previous section in terms of the two above parametrizations keeping photons as the only gauge fields.
In the case of the exponential parametrization we have the $\cO(p^2)$ Lagrangian,
\bear
{\cal L}_2(\pi,h,\gamma)  & = & \frac{1}{2}\partial_\mu  h \partial^{\mu}h+\frac{1}{2}
  \mF(h)       
\bigg[\frac{v^2}{\pi^2}\sin^2\frac{\pi}{v}
\bigg(\delta_{ab}-\frac{\pi^a\pi^b}{\pi^2}\bigg)+\frac{\pi^a\pi^b}{\pi^2}\bigg]
\partial_{\mu}\pi^a\partial^{\mu}\pi^b        \nonumber    \\
    &  & \hspace*{-1cm}+
\bigg\{  i e    \mF(h)     
A^\mu \bigg[\frac{v^2}{\pi^2}\sin^2\frac{\pi}{v}\partial_\mu \pi^+\pi^-
  +   \frac{v}{2} \frac{\pi^+\pi^-}{\pi^3}\partial_\mu\pi^2\sin\frac{\pi}{v}
  \bigg(\cos\frac{\pi}{v}-\frac{\pi}{v}\sin\frac{\pi}{v}\bigg) \bigg]
   + h.c. \bigg\}    \nonumber  \\
  &  & \hspace*{-1cm}+  e^2
  \mF(h)         
A_\mu A^\mu\pi^+\pi^- \frac{v^2}{\pi^2}\sin^2\frac{\pi}{v},
\eear
and for the spherical parametrization,
\bear
{\cal L}_2(\omega,h,\gamma)    & = & \frac{1}{2}\partial_\mu  h \partial^{\mu}h+\frac{1}{2}
\mF(h)       
\left(\delta_{ab}+\frac{\omega^a\omega^b}{v^2-\omega^2}\right)
\partial_{\mu}\omega^a\partial^{\mu}\omega^b        \nonumber    \\
    & +  &   i e
  \mF(h)         
A^\mu    (\partial_\mu \omega^+\omega^-  -  \omega^+\partial_\mu\omega^- )  + e^2
  \mF(h)        
A_\mu   A^\mu  \omega^+\omega^-
\eear
where in both cases, the relevant terms in $\mF(h)$ are:
\be
  \mF(h)          
= 1 + 2 a \frac{h}{v}+ b \frac{h^2}{v^2}.
\ee
These  second coordinates give rise to a much simpler Lagrangian and simpler Feynman rules and diagrams for a given processes. For example
using the $\omega$ coordinates photons only couple to two WBGB. On the other hand the exponential parametrization produces couplings of photons to an arbitrary large number of WBGBs.

For the two processes of interest here, $\gamma \gamma \rightarrow z z $ and  $\gamma \gamma \rightarrow w^+ w^- $, the relevant
 $\cO(p^2)$  Lagrangians   for the two parametrizations considered here are:
\bear
 {\cal L}_2(\pi,h,\gamma)    & = & \frac{1}{2}\partial_\mu  h \partial^{\mu}h    + \frac{1}{2}
\mF(h)        
(2\partial_\mu\pi^+   \partial^{\mu}\pi^- + \partial_\mu\pi^0   \partial^{\mu}\pi^0 )     \nonumber \\
& +  & \frac{       \mF(h)      
}{6 v^2}[(\partial_\mu \pi^+  \pi^- + \pi^+ \partial_\mu \pi^-  +\pi^0\partial_\mu \pi^0)^2
-\pi^2(2\partial_\mu\pi^+   \partial^{\mu}\pi^- + \partial_\mu\pi^0   \partial^{\mu}\pi^0 )  ]
\nonumber \\
   & + &
  i e  \mF(h)  \bigg\{          
A^\mu\bigg[(\partial_\mu \pi^+\pi^-\bigg(1-\frac{\pi^2}{3 v^2}\bigg)-\frac{\pi^+\pi^-}{6v^2}\partial_\mu\pi^2  \bigg]
   +  h.c.  \bigg\}
   \nonumber \\
    &+ & e^2
   \mF(h)        
A_\mu A^\mu  \pi^+\pi^-\bigg(1-\frac{\pi^2}{3 v^2}\bigg)
\eear
and
\bear
{\cal L}_2(\omega,h,\gamma)     & = & \frac{1}{2}\partial_\mu  h \partial^{\mu}h+
\frac{1}{2}
   \mF(h)      
(2\partial_\mu\omega^+   \partial^{\mu}\omega^- + \partial_\mu\omega^0   \partial^{\mu}\omega^0 )     \nonumber \\
& +  & \frac{      \mF(h)    
}{2 v^2}(\partial_\mu\omega^+\omega^- +\omega^+\partial_\mu\omega^- +
\omega^0  \partial_\mu \omega^ 0)^2   \nonumber  \\
   & +  &   i e
     \mF(h)      
A^\mu (\partial_\mu \omega^+\omega^-  -  \omega^+\partial_\mu\omega^- ) + e^2
    \mF(h)      
A_\mu A^\mu \omega^+\omega^- .
\eear
 These are the relevant pieces of the  $\mL_2$ Lagrangian
(in the corresponding coset parametrization)  for our photon-photon computation,
where  we use the Landau gauge and the WBGBs are massless
   and      we are, in practice, taking $m_h=0$.
All these assumptions together with the fact that we are using dimensional regularization
will reduce considerably the number of diagrams to be computed.

{}From these Lagrangians above it is now straightforward to get the Feynman rules for the relevant couplings. In Appendix A we show the relevant rules obtained from both parametrizations for incoming particles.
As one could have expected, those rules involving less than four WBGBs are  exactly the same in both parametrizations. This is because  both coordinates differ in terms which are at least quadratic  in the WBGBs. Thus the vertices with four WBGBs are indeed different in both parametrizations if the  WBGBs are off-shell but they coincide for on-shell legs to guarantee that the corresponding $S$ matrix elements are parametrization independent. Notice that in particular the vertex with four neutral WBGBs $zzzz$ vanishes in the exponential coordinates case  but it  does not vanish in the spherical ones (off-shell). Therefore, this vertex will contribute to one-loop diagrams in one case but not in the other one. This shows that
the independence of the $S$ matrix elements on the particular parametrization at the one-loop level is not a trivial result at all (see \cite{mybook} and references therein).
This fact was observed for the first time in~\cite{DobadoMorales} in the context of $SU(2)$ ChPT for the process
$\gamma \gamma \rightarrow \pi^0\pi^0$ to one loop and in the chiral limit. In this case the on-shell one-loop amplitude
requires the computation of four diagrams using the exponential coordinates
(see for example~\cite{Donoghue})
but only two diagrams when one uses the spherical parametrization.

Besides, the vertices with four WBGBs and one photon are also different in both
par\-am\-etrizations,
even for on-shell legs but this is not a problem since the corresponding physical process ($S$ matrix elements) have other contributions at the tree level.

\section{
Analytical results for the one-loop  $\gamma\gamma\to w^a w^b$  scattering amplitudes }
\label{sec.results}

Following the procedure described in detail in the previous sections, we proceed now to perform
the computation of the $\gamma\gamma\to w^a w^b$ amplitudes up to NLO in the chiral expansion.
In the charge basis for the WBGBs, this corresponds to the computation
of the $\gamma\gamma \to z z$ and $\gamma\gamma\to w^+w^-$ amplitudes, respectively
related to $\gamma\gamma\to Z_L Z_L$ and $\gamma\gamma\to W^+_L W^-_L$ amplitudes through
the Equivalence Theorem.

In addition, under the assumptions and approximations taken in the present article,
many Feynman diagrams suffer
important simplifications. First, we are left just with the one-loop diagrams with only scalar fields, $h$, $w^{\pm}$ and $z$,
in the internal lines. Second, in the massless case considered here, namely, in the Landau gauge with WBGB's poles at zero
and negligible Higgs mass, $m_h\ll\sqrt{s}\ll 4\pi v$, all the one point Feynman integrals
$A_0(m^2)$ with $m= m_{h,w^\pm,z}$
   vanish
in dimensional regularization.
Since this type of diagrams do not contribute in our case
they have not been included in our plots together with the remaining
one-loop graphs for $\gamma \gamma\to z z$
and $\gamma\gamma\to w^+ w^-$.  Likewise, there is no wave-function
renormalization in the massless case ($Z=1$) and there is no need for
a separate study of the self-energies.
Nonetheless, we would like to emphasize that the three eliminated contributions (Feynman diagrams with one-point scalar Feynman integrals, diagrams with internal EW gauge bosons and
the wave-functions renormalization) must be included if the calculation
is required to account also for subleading corrections, like
{    ${\cal O} (m_{W,Z,h}^2/(16\pi^2 v^2))$     }
corrections.~\footnote{
 For instance, if we had considered $m_h\neq 0$ in our analysis the WBGB
wave-function renormalization would have received corrections as
$Z=1 +\frac{(b-a^2)}{16\pi^2 v^2} A_0(m_h^2)-\frac{a^2 m_h^2}{32\pi^2 v^2}$~\cite{Espriu:2013B},
with $A_0(m_h^2)$ the one-point Feynman integral.}

The present computation has been performed with the help of FeynRules, FeynArts and FormCalc~\cite{FAFC} and has also been double-checked
through two independent computations, one in the
exponential parametrization of $U(x)$ and the other with $U(x)$ expressed in spherical coordinates.
We found full agreement in the final results from both parametrizations when the two external WBGBs and the two external photons were set on-shell,
as expected~\cite{mybook}.

We present the results for the scattering amplitudes by using the compact form that is determined by electromagnetic gauge invariance. Thus, the results for the $\gamma(k_1,\epsilon_1)\gamma(k_2,\epsilon_2)\to w^+(p_1) w^-(p_2)$ and $\gamma(k_1,\epsilon_1)\gamma(k_2,\epsilon_2)\to z(p_1) z(p_2)$ amplitudes will be given (as in the analogous case of scattering of photons into pions)
by the general  decomposition~\cite{DobadoMorales,Morales:92,Dawson:91,Bijnens:1988,Burgi:1996}

\be
\mM   =
ie^2 (\epsilon_1^\mu \epsilon_2^\nu T_{\mu\nu}^{(1)}) A(s,t,u)+
ie^2 (\epsilon_1^\mu \epsilon_2^\nu T_{\mu\nu}^{(2)})B(s,t,u),
\ee
that is written in terms of the two independent Lorentz structures,
\bear
(\epsilon_1^\mu\epsilon_2^\nu T^{(1)}_{\mu\nu}) &=&
\frac{s}{2} (\epsilon_1 \epsilon_2) - (\epsilon_1 k_2)(\epsilon_2 k_1),
\label{eq.T1-def}
\\
(\epsilon_1^\mu\epsilon_2^\nu T^{(2)}_{\mu\nu}) &=&
2 s (\epsilon_1 \Delta)(\epsilon_2 \Delta)
- (t-u)^2 (\epsilon_1\epsilon_2)
- 2 (t-u) [(\epsilon_1\Delta)(\epsilon_2 k_1) - (\epsilon_1 k_2) (\epsilon_2 \Delta)],
\nn\\
\label{eq.T2-def}
\eear
with the Mandelstam variables defined as usual, $s=(p_1+p_2)^2$, $t=(k_1-p_1)^2$
and $u=(k_1-p_2)^2$, the relevant momentum combination is defined as $\Delta^\mu\equiv p_1^\mu -p_2^\mu$,
and the $\epsilon_i$'s are the  polarization vectors of the external photons.

As for the separate contributions to the scattering amplitudes we will follow the steps and notation that we have introduced in section 3. Thus, the results for the total amplitudes will be reported into two parts, correspondingly to the explained LO and NLO contributions:
\be
\mM =\mM_{\rm LO}+\mM_{\rm NLO}{\rm , and}\; A=A_{\rm LO}+A_{\rm NLO},\;B=B_{\rm LO}+B_{\rm NLO}
\ee
$\mM_{\rm LO}$ involves the computation of the tree graphs from ${\cal L}_2$ which are indeed none for the
$\gamma\gamma\to zz$ case and those illustrated in Fig.~3a for the $\gamma\gamma\to  w^+w^-$ case. $\mM_{\rm NLO}$ includes the two contributions: from the one-loop diagrams and from the tree graphs, as explained in Eq.(\ref{NLO}). The tree graphs contributing to $\mM_{\rm NLO}$ are displayed in
Fig.~1 for $\gamma\gamma\to zz$ and in~Fig.~3b for $\gamma\gamma\to  w^+w^-$. Finally, the contributing one-loop diagrams are displayed
in~Fig.~2 for $\gamma\gamma\to zz$ and in~Fig.~4 for  $\gamma\gamma\to  w^+w^-$.
Although, for shortness, we report in the following on the total result for the sum of all the one-loop diagrams, i.e,
$\mM^{\rm 1-loop}(\gamma\gamma\to zz)=\sum_{i=1}^{10}  \mM^i(\gamma\gamma\to zz)$
and $\mM^{\rm 1-loop}(\gamma\gamma\to w^+w^-)=\sum_{i=1}^{39}  \mM^i(\gamma\gamma\to w^+w^-)$,
the individual analytical results for each diagram are also
provided, for completeness, in the
  Appendix~\ref{app.diagrams}.

Before presenting the final results, there are some subtleties/curiosities in the calculation of the loop diagrams that we find interesting to comment:
\begin{itemize}

\item{} The diagrams~5 and 7 in $\gamma\gamma\to zz$ (Fig.~\ref{ggtozz})
and 26 and 27 in $\gamma\gamma\to w^+w^-$ (Fig.~\ref{ggtoww})
are always absent in the spherical parametrization
of $U(x)$, even off-shell, as there is no vertex with one photon and four WBGBs  in these
coset coordinates.
In the exponential parametrization this difference  is compensated by the diagrams which contain
vertices with four WBGBs (diagrams~1, 2 and
  6
in Fig.~\ref{ggtozz}  and 1, 2, 16, 28, 29, 30 and 31 in Fig.~\ref{ggtoww},
in addition to some diagrams with one-point Feynman integrals not plotted therein),
which are different in
spherical and exponential coordinates.

Nevertheless,  in the Landau gauge considered here, where the WBGBs  are massless,
the diagrams~5 and 7 in  Fig.~\ref{ggtozz}
and 26 and 27 in  Fig.~\ref{ggtoww}
happen to be  zero in both  parametrizations of $U(x)$. Likewise,
diagrams~1, 2 and
   6
in Fig.~\ref{ggtozz} and 1, 2, 16, 28, 29, 30 and 31 in Fig.~\ref{ggtoww} turn out to be respectively identical in
the two coset coordinates.

\item In the Landau gauge, the diagrams~8 and 9 in $\gamma\gamma\to zz$ (Fig.~\ref{ggtozz})
and 28--35 in $\gamma\gamma\to w^+w^-$ (Fig.~\ref{ggtoww})
are zero in both parametrizations, exponential and spherical.

\item For the Landau gauge and for negligible Higgs mass,
the diagrams~36--39 in $\gamma\gamma\to w^+w^-$ (Fig.~\ref{ggtoww})
are zero in both parametrizations, exponential and spherical.

\end{itemize}

\subsection{Analytical results for $\gamma\gamma\to zz$}

\begin{figure}[!t]
\centering
\begin{center}
\begin{minipage}[b]{0.4\textwidth}
\centering
\psfrag{X}{${\rm \gamma}$}
\includegraphics[width=3.5cm]{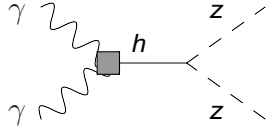}
\end{minipage}
\caption{{\small Single tree-level diagram contributing to $\gamma\gamma\to zz$
   at  $\cO(e^2 p^2)$    
The vertex with a shaded box stands for
 an interaction from $\mL_4$   and the normal vertex  is for one from  $\mL_2$.
}}
\label{fig.tree-ggtozz}
\end{center}
\end{figure}

\begin{figure}[!t]
\begin{center}
\begin{tabular}{c c c c c}
\psfig{file=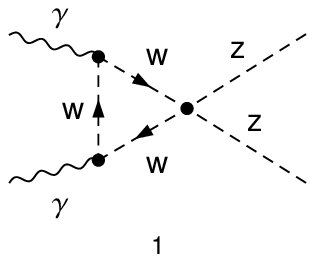,width=2.5cm,clip=} & \psfig{file=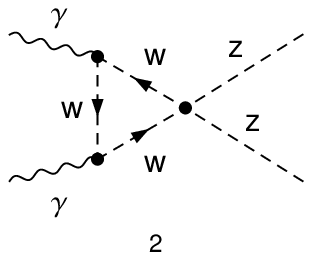,width=2.5cm,clip=}
& \psfig{file=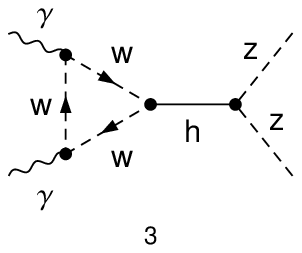,width=2.5cm,clip=} & \psfig{file=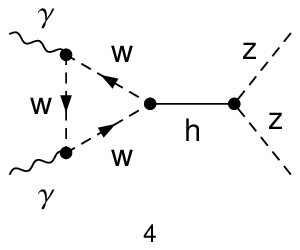,width=2.5cm,clip=}
& \psfig{file=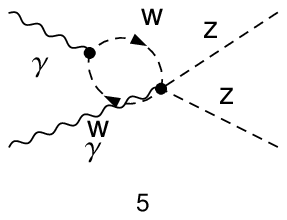,width=2.5cm,clip=}
\\
\psfig{file=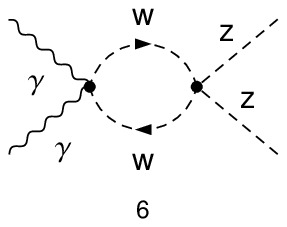,width=2.5cm,clip=} & \psfig{file=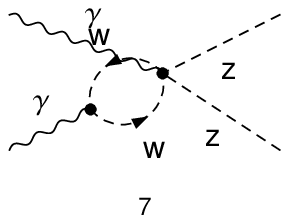,width=2.5cm,clip=}
& \psfig{file=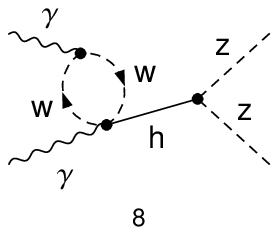,width=2.5cm,clip=} & \psfig{file=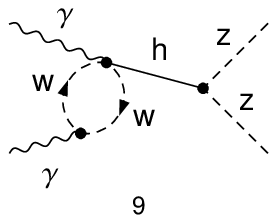,width=2.5cm,clip=}
& \psfig{file=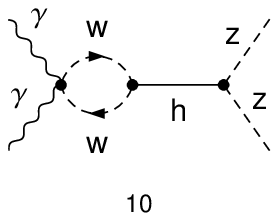,width=2.5cm,clip=}
\end{tabular}
\caption{{\small  One-loop diagrams for $\gamma\gamma\to zz$. }}
\label{ggtozz}
\end{center}
\end{figure}

The $\gamma\gamma$ scattering amplitude into a pair of neutral WBGBs is found to be zero at lowest order
in the chiral expansion i.e. at   $\cO(e^2)$ (in agreement with ~\cite{Morales:92} and also with the analogous scattering amplitudes for the pions case~\cite{Donoghue,Dawson:91}):
\be
\mM(\gamma \gamma \to zz)_{\rm LO}=0 \, .
\label{LOzz}
\ee
At NLO, i.e at  $\cO(e^2p^2)$      
one has one-loop and tree-level contributions to $A(s,t,u)$ and $B(s,t,u)$,
which depend only on the kinematical variable $s$
at this order.  After adding all the contributions we find the following extremely simple result:
\bear
A(\gamma \gamma \to zz)_{\rm NLO}&=&
\Frac{2 a c_\gamma^r}{v^2} + \Frac{ (a^2-1)}{4\pi^2v^2},
\label{eq.A-zz} \\
B(\gamma \gamma \to zz)_{\rm NLO}&=& 0,
\label{eq.B-zz}
\eear
where the term proportional to $c^r_\gamma$ comes from the tree-level
 $\cO(e^2 p^2)$  contributions        
(Fig.~\ref{fig.tree-ggtozz})
and the term proportional to $(a^2-1)$ comes from the one-loop diagrams (Fig.~\ref{ggtozz}).
Independent diagrams (e.g. diagram~6) are  in general divergent.
However, in dimensional regularization,
the final one-loop amplitude turns out to be UV finite  when all the contributions are put together. Therefore the $\mL_4$ chiral
  parameter
$c^r_\gamma$ does not need to be renormalized to cancel  the UV-divergences and, in consequence,
\begin{equation}
c^r_\gamma=c_\gamma.
\label{renormcgamma}
\end{equation}
Notice also that by setting $a=c_\gamma =0$ in our formulas above we recover exactly the result found in ~\cite{Morales:92} for the case of Higgsless ECL, which in turn agreed with the analogous result for the amplitude in the pions case~\cite{Donoghue,Dawson:91}.

\subsection{Analytical results for $\gamma\gamma\to w^+ w^-$}


\begin{figure}[!t]
\centering
\begin{center}
\begin{minipage}[b]{0.4\textwidth}
\centering
\psfrag{X}{$\gamma$}
\includegraphics[width=5.5cm]{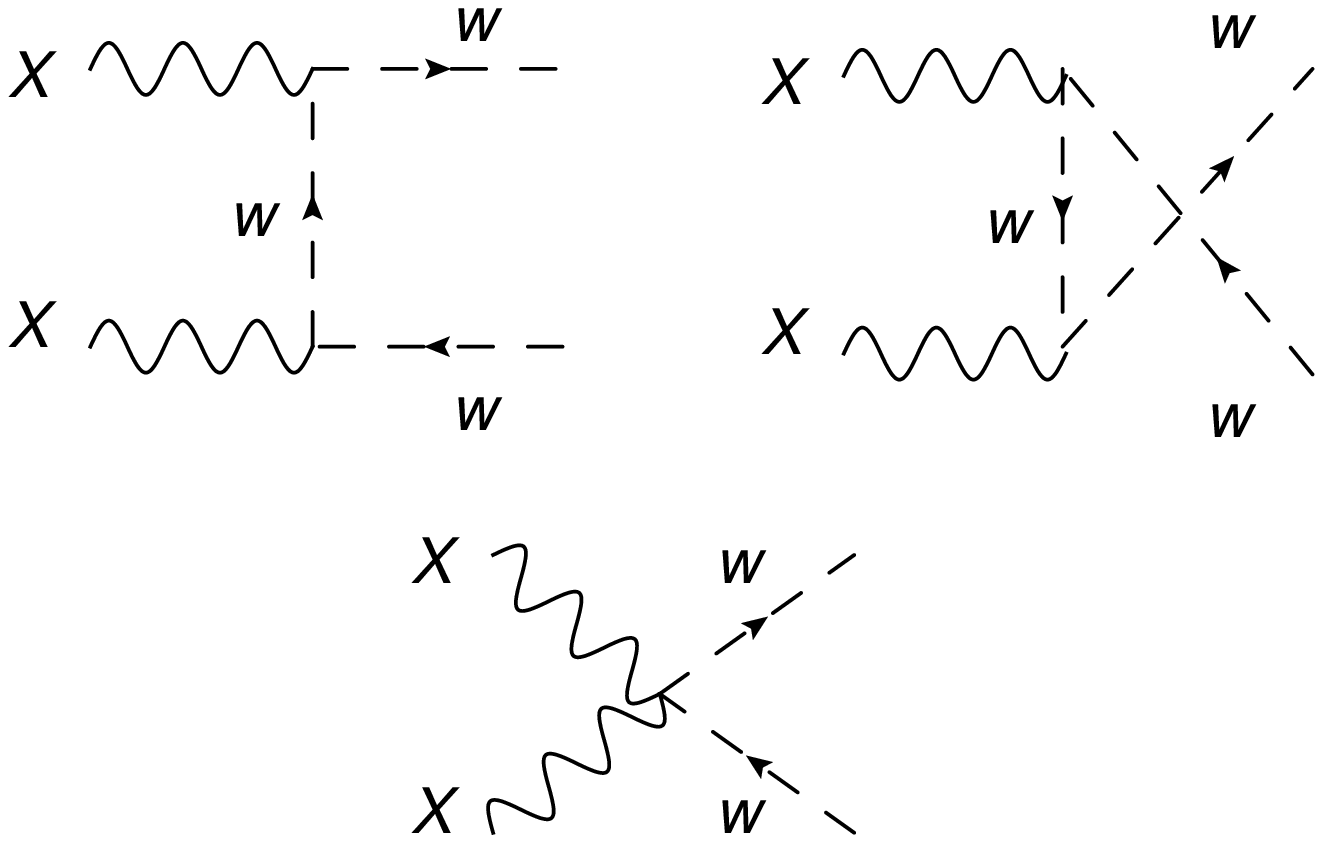}
\\
\vspace*{1.25cm}
{\bf a)}
\end{minipage}
\hspace*{0.25cm}
\begin{minipage}[b]{0.4\textwidth}
\centering
\psfrag{X}{$\gamma$}
\includegraphics[width=5cm]{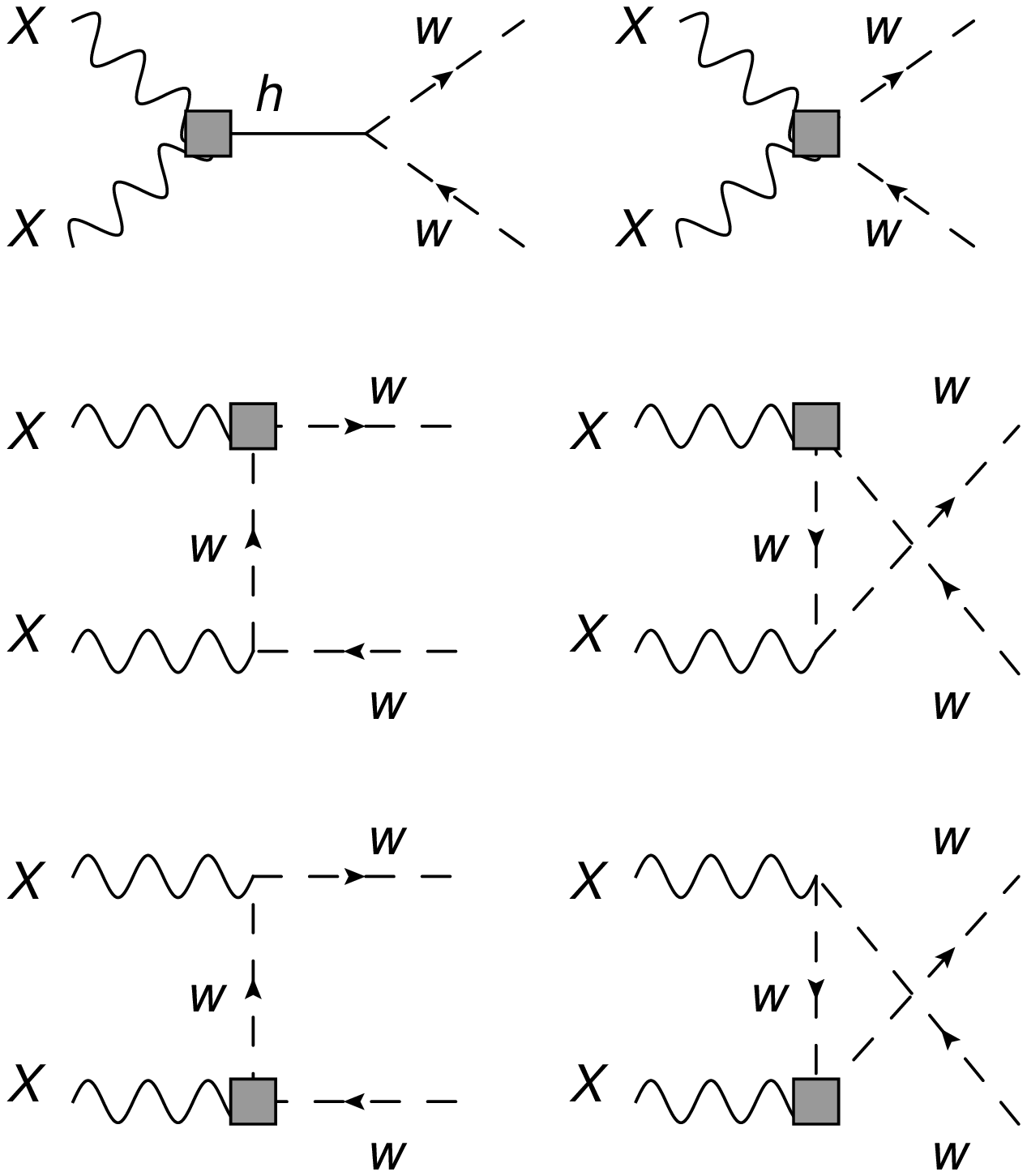}
\\
{\bf b)}
\end{minipage}
\caption{{\small Tree-level diagrams for $\gamma\gamma\to w^+ w^-$
   at    $\cO(e^2)$                 
{\bf (a)} and    $\cO(e^2 p^2)$         
{\bf (b)}.
}}
\label{fig.tree-ggtoww}
\end{center}
\end{figure}

\begin{figure}[!t]
\begin{center}
\begin{tabular}{c c c c c}
\psfig{file=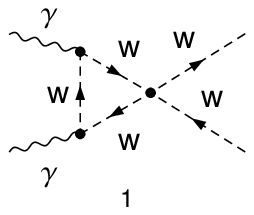,width=2.4cm,clip=} & \psfig{file=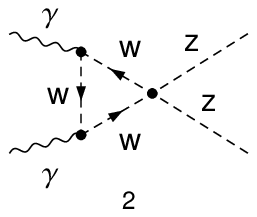,width=2.4cm,clip=}
& \psfig{file=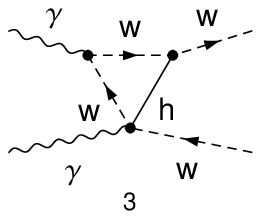,width=2.4cm,clip=} & \psfig{file=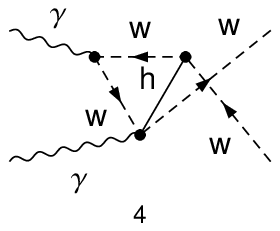,width=2.4cm,clip=}
& \psfig{file=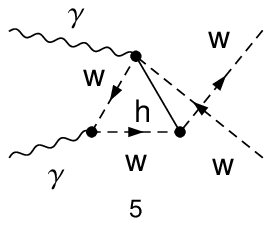,width=2.4cm,clip=}
\\
\psfig{file=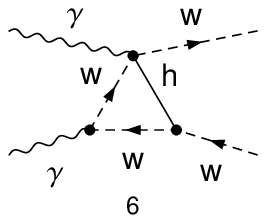,width=2.4cm,clip=} & \psfig{file=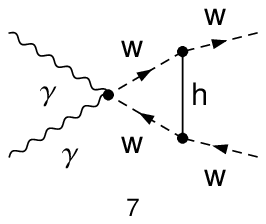,width=2.4cm,clip=}
& \psfig{file=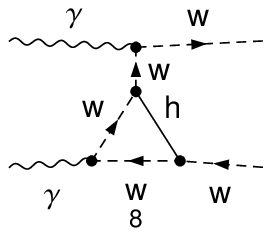,width=2.4cm,clip=} & \psfig{file=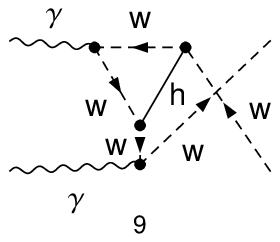,width=2.4cm,clip=}
& \psfig{file=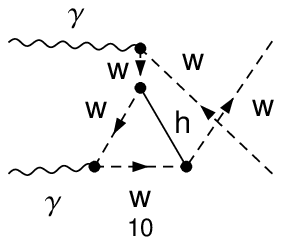,width=2.4cm,clip=}
\\
\psfig{file=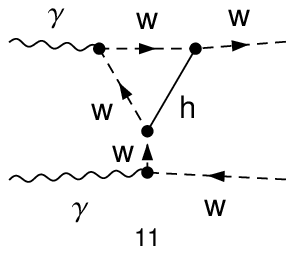,width=2.4cm,clip=} & \psfig{file=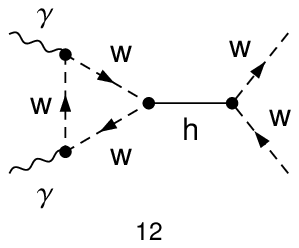,width=2.4cm,clip=}
& \psfig{file=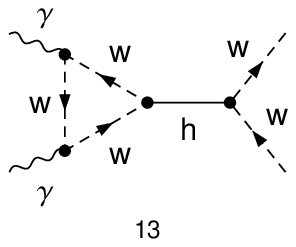,width=2.4cm,clip=} & \psfig{file=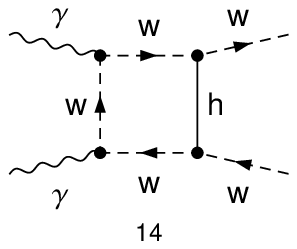,width=2.4cm,clip=}
& \psfig{file=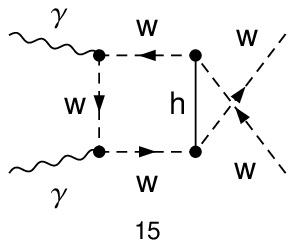,width=2.4cm,clip=}
\\
\psfig{file=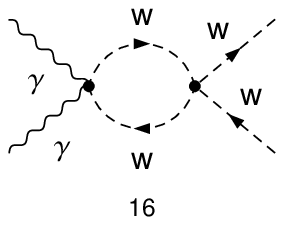,width=2.4cm,clip=} & \psfig{file=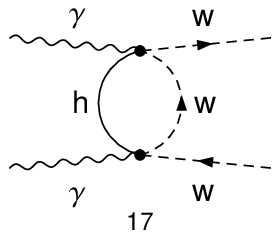,width=2.4cm,clip=}
& \psfig{file=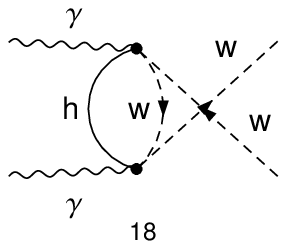,width=2.4cm,clip=} & \psfig{file=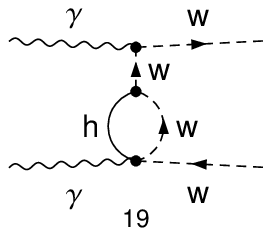,width=2.4cm,clip=}
& \psfig{file=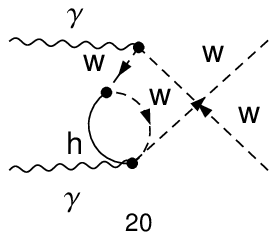,width=2.4cm,clip=}
\\
\psfig{file=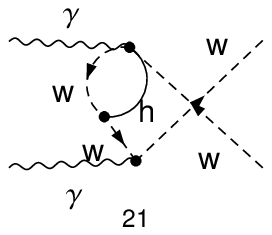,width=2.4cm,clip=} & \psfig{file=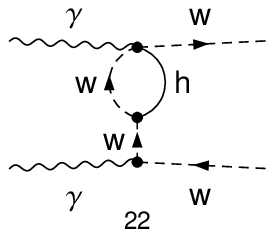,width=2.4cm,clip=}
& \psfig{file=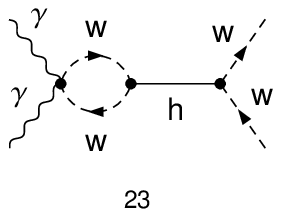,width=2.4cm,clip=} & \psfig{file=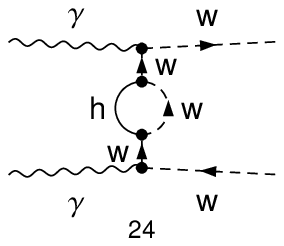,width=2.4cm,clip=}
& \psfig{file=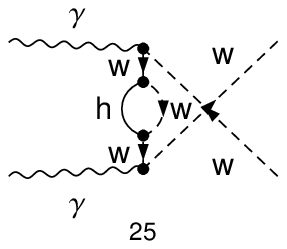,width=2.4cm,clip=}
\\
\psfig{file=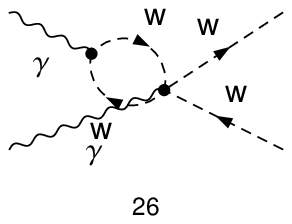,width=2.4cm,clip=} & \psfig{file=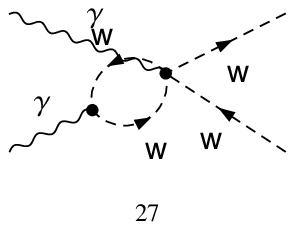,width=2.4cm,clip=}
& \psfig{file=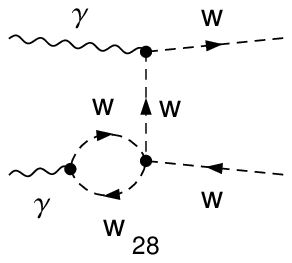,width=2.4cm,clip=} & \psfig{file=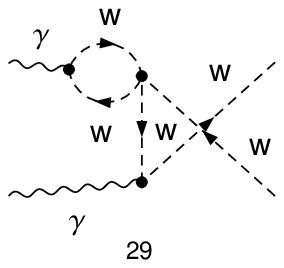,width=2.4cm,clip=}
& \psfig{file=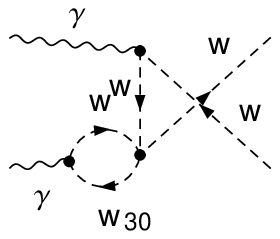,width=2.4cm,clip=}
\\
\psfig{file=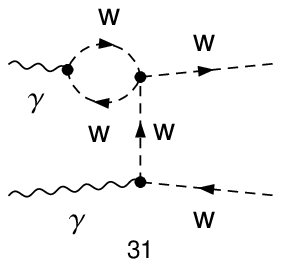,width=2.4cm,clip=} & \psfig{file=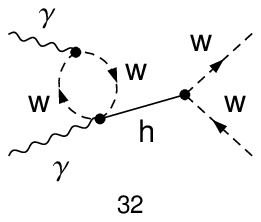,width=2.4cm,clip=}
& \psfig{file=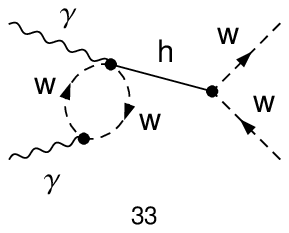,width=2.4cm,clip=} & \psfig{file=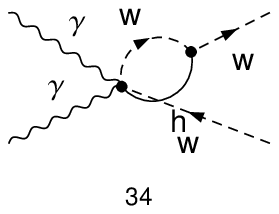,width=2.4cm,clip=}
& \psfig{file=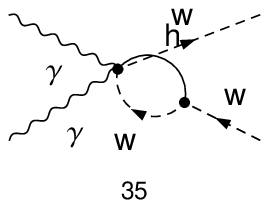,width=2.4cm,clip=}
\\
\psfig{file=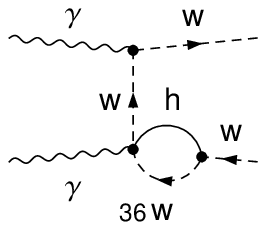,width=2.4cm,clip=} & \psfig{file=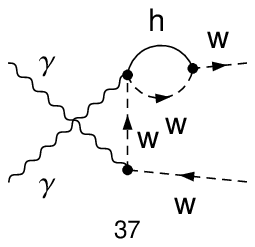,width=2.4cm,clip=}
& \psfig{file=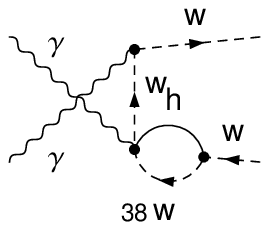,width=2.4cm,clip=} & \psfig{file=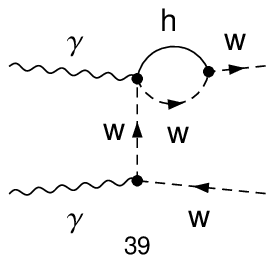,width=2.4cm,clip=}
&
\end{tabular}
\caption{{\small One-loop diagrams for $\gamma\gamma\to w^+ w^-$. }}
\label{ggtoww}
\end{center}
\end{figure}

At LO, i.e. at $\cO(e^2)$, the contributing diagrams are displayed in Fig.~\ref{fig.tree-ggtoww}a. We find the following result (in agreement with ~\cite{Morales:92} and also with the analogous scattering amplitude for the pions case~\cite{Bijnens:1988}):
\be
A(\gamma\gamma\to w^+ w^-)_{\rm LO}    
= 2 s B(\gamma\gamma\to w^+ w^-)_{\rm  LO}    
= -\Frac{1}{t} - \Frac{1}{u},
\ee
where one can observe the contributions from the $t$ and $u$--channel $w^+$ exchanges.

For the NLO contributions,\ie those of $\cO(e^2 p^2)$,       
we obtain again an extremely simple result after
combining all the various tree-level (Fig.~\ref{fig.tree-ggtoww}) and
one-loop diagrams (Fig.~\ref{ggtoww}) (see the Appendix B for the separate contributions from each diagram):
\bear
A(\gamma\gamma\to w^+ w^-)_{\rm NLO} &=&
\Frac{8(a^r_1-a^r_2+a^r_3)}{v^2} + \Frac{2 a c_{\gamma}^{r}}{v^2} + \Frac{(a^2-1)}{8\pi^2v^2},
\label{eq.A-ww}
\\
B(\gamma\gamma\to w^+ w^-)_{\rm NLO} &=& 0.
\label{eq.B-ww}
\eear
While $B$ does not suffer corrections
  of  $\cO(e^2 p^2)$,         
one has that at this order        
there are tree-level contributions to $A$ and these are
proportional to the combination of parameters $(a_1-a_2+a_3)$ and to $ac_\gamma$.
The total one-loop contribution is given by the term $(a^2-1)$ in the right-hand side
in Eq.~\eqref{eq.A-ww}.
Surprisingly, we find that again the one-loop UV divergences exactly cancel out when
all diagrams are put together and, in consequence, no renormalization of the combination of
   $\mL_4$         
chiral parameters in~\eqref{eq.A-ww} is required.
Therefore, from Eqs.~\eqref{eq.A-ww} and~\eqref{eq.B-ww}
and using Eq.~\eqref{renormcgamma}, we find:
\be
(a_1^r-a_2^r+a_3^r)=(a_1-a_2+a_3)
\label{invariant}
\ee
This result is highly non-trivial and comes after subtle cancelations of the various contributions.
For instance, the box diagrams~14 and 15 yield  a complicate Lorentz structure
and depend on the scalar two-point Feynman integrals $B_0(s,0,0)$, $B_0(t,0,0)$
and $B_0(u,0,0)$. See
   Appendix~\ref{app.diagrams}
for more details. Finally, notice also that,   as in the previous scattering
process, if we set $a=c_\gamma=0$ we recover exactly the result found in ~\cite{Morales:92} for the case of Higgsless ECL, which in turn agreed with the analogous result for the amplitude in the pions case~\cite{Bijnens:1988}.

\section{Discussion}
\label{sec.discusion}
First of all, we would like to remark again that our results for the one-loop scattering amplitudes $\mM(\gamma \gamma \to w^+w^-)$ and $\mM(\gamma \gamma \to zz)$, presented in the previous sections, converge to the corresponding results of the Higgsless ECL case in \cite{Morales:92} if we set properly the Higgs-like parameters, namely, if we set $a=c_\gamma=0$. In particular, it is interesting to notice that the combination of EW chiral parameters $a_i$ that enters in $\gamma \gamma \to w^+w^-$, given in Eq.~\eqref{invariant}, which we have found to be renormalization group invariant, is the same in both cases, the ECLh and ECL. This finding, apart of being a convenient check of our computation, it is by itself a quite interesting result, since the dynamical Higgs boson is contributing non-trivially in the loop diagrams of the present ECLh case and therefore it is contributing to the renormalization of each of the $a_i$'s. In contrast, the Higgs field is totaly absent in the
ECL case. The fact that we have found the same renormalization group invariant combination $(a_1-a_2+a_3)$ for this scattering process in the ECLh case as in the ECL could be just a coincidence for this particular case, or it could be a more general result. In other words, one can wonder if there are other renormalization group invariant combinations of the $a_i's$ that are common to the ECLh and ECL cases and if there is any fundamental explanation for this. Obviously to give a complete answer to this question one should compute the full one-loop action (in both the ECLh and the ECL) and set the proper renormalization of all the parameters involved in these chiral Lagrangians, but this is clearly beyond the scope of this work.

Secondly, we would like to point out some other interesting aspects that are suggested by the simple final formulas that we have found in the ECLh case for the $\mM(\gamma \gamma \to w^+w^-)$ and $\mM(\gamma \gamma \to zz)$ one-loop amplitudes.
We believe that this simplicity of the final results seems to be hinting some important underlying features in these ECLh models.
A possible explanation could be the existence of a more appropriate choice of
the degrees of freedom describing the WBGBs and the Higgs boson together in the massless Higgs
limit, which could point towards a larger symmetry, as it is indeed the case of the $SO(5)/SO(4)$ model. As commented above this massless Higgs limit is appropriate at high energies where the Equivalence Theorem
applies because of the phenomenological fact that the boson masses are relatively light and close to each other, $m_h \sim m_W \sim m_Z \sim$ ${\cal O}(100\, {\rm GeV})$. For illustration and comparison with the present ECLh case,
in appendix~\ref{app.MCHM} we have computed the
one-loop  amplitude, for $\gamma \gamma \to w^a w^b$ scattering, in the context of the
$SO(5)/SO(4)$ model. There it is shown that, by using an appropriate parametrization
of the $S^4$ coset relevant for these models, which reduces the number of contributing one-loop diagrams drastically, the computation can be greatly simplified and, indeed, we get the same result as we got previously for
 $\mM(\gamma \gamma \to zz)^{\rm 1-loop}$ and $\mM(\gamma \gamma \to w^+ w^-)^{\rm 1-loop}$
after a tedious calculation and by setting the parameter $a$ to the corresponding value
  in the  $SO(5)/SO(4)$ MCHM (see Eq.\eqref{models}).
In this way, we understand the simplicity of the results because the processes considered here are independent of
the $b$ parameter at the one-loop level, thus making the prediction
   of  the $SO(5)/SO(4)$ MCHM
for the $\mM(\gamma \gamma \to w^a w^b)^{\rm 1-loop}$ scattering amplitudes to be a
universal prediction.

Furthermore, in the computation of the previous section, one can see that the loop suppression
in the full amplitudes
is actually stronger than that provided by naive dimensional analysis,
as we find strong cancellations between diagrams such that the suppression of the loop contributions
is not the usual in chiral Lagrangians,   $\cO\left(E^2/(16\pi^2 v^2)\right)$, with $E^2=s,t,u$,
but rather  $\cO\left((1-a^2) E^2/(16\pi^2 v^2)\right)$.
 This  can be immediately understood thanks to the computation
 in the $SO(5)/SO(4)$ context
(appendix~\ref{app.MCHM}), where there are just two contributing one-loop
   topologies  (see Fig.~\ref{fig.MCHM})
and  each of them is suppressed  by
$\left(E^2/(16\pi^2 f^2)\right)$,  with
${  \Lambda_{\rm ECLh}\sim 4\pi f > 4\pi v   }$ the true characteristic cut-off scale of the
ECLh.

Finally, motivated by a future phenomenological analysis of our results presented here, we propose to study these scattering $\gamma \gamma \to w^+w^-$ and $\gamma \gamma \to zz$ processes together with other observables that involve the same subset of chiral parameters, such that one can perform
in the future a global analysis of all these observables together, compare them with data, and get useful information from this analysis
 on the values  preferred by data  for    these chiral parameters.
With this purpose in mind, we have considered a set
of four additional observables:   the $h\to\gamma \gamma$ decay width,
the EW precision $S$ parameter, the $\gamma^*\to w^+w^-$ vector form-factor and
the $\gamma^* \gamma\to h$ transition form-factor, whose detailed formulas are collected in Appendix~\ref{app.related-obs}.
As one can see in Table~\ref{tab.relevant-couplings},
putting everything together,  the whole system of six observables is over-constrained. By means of these additional four
appropriate  observables it
is possible to fix the four relevant combinations of chiral parameters in the
considered amplitudes, for instance, $a$, $c_\gamma^r$, $a_1^r$ and $(a_2^r-a_3^r)$,
and then predict from them the remaining ones. Besides,
we find that even though  $c_\gamma^r$ and the combination $(a_1^r-a_2^r+a_3^r)$
are renormalization group invariant, if one looks into the separate contributions, the parameter $a_1$ and the combination
$(a_2-a_3)$  need to be renormalized.

One can also learn from our study about the specific running of the involved chiral parameters.
Generically, the relation between a given renormalized chiral parameter $C^r(\mu)$ and the corresponding bare
parameter $C^{(B)}$ from the $\mL_4$ Lagrangian
(e.g. $a_1$) is given by
\bear
  C^r(\mu) &=& C^{(B)} + \Frac{\Gamma_C}{32\pi^2}  \Frac{1}{\hat{\epsilon}},
\eear
where we have performed the $\overline{MS}$ subtraction of the  UV divergence
  $1/\hat{\epsilon}$ defined in Eq.~\eqref{epsilon},
with $D=4-2\epsilon$.  The running of the renormalized couplings are, in consequence, given by:
\bear
\Frac{dC^r}{d\ln{\mu}} &=& -\Frac{\Gamma_C}{16\pi^2}.
\eear

The relevant $\mL_4$        
parameters in our $\gamma\gamma\to w^a w^b$
analysis are $C=a_1,\, a_2,\, a_3, \, c_\gamma$   and their running
  is
shown in Table~\ref{tab.running}.
One can see that $c_\gamma^r$ and the combination ${a_1^r-a_2^r+a_3^r}$
are  renormalization group invariant.
   The latter combination is renormalization group invariant,
as it happened in the case of the Higgsless Electroweak Chiral Lagrangian (ECL)~\cite{Morales:94}.
This is also trivially true in the SM, where the $c_\gamma$ and the $a_i$ are absent.
Indeed, since $a=b=1$ for all the linear models where the Higgs
is introduced through a complex doublet $\Phi$~\cite{linear-EFT-Manohar,SILH}, the renormalized couplings $c_\gamma^r$,
$a_1^r$ and  the combination $(a_2^r-a_3^r)$ do not run in those cases.
  Our result for the running of $(a_2^r-a_3^r)$  in Table~\ref{tab.running} also agrees with the QCD determination
for the analogous chiral parameter in Ref.~\cite{Talavera:2014}.

In addition, although they do not play any role in the present article, we have also included for completeness in this table the running of
$a_4^r$ and $a_5^r$. These two chiral parameters enter in the  $W^+W^-$, $ZZ$ and $hh$ scattering and therefore they will play a very relevant role in the future analysis of LHC data at $\sqrt{s}=13\, {\rm TeV}$. Their running
have been recently determined in the one-loop analyses from
  Refs.~\cite{Espriu:2013B, Dobado:2013}.   We
have also included them, for completeness, in the last two rows of Table~\ref{tab.running}.

Regarding the second column in Table~\ref{tab.relevant-couplings}, we wish to emphasize once again that the parameters of the  $\mL_2$ Lagrangian ($a$ in this case)
do not get renormalized in dimensional regularization, as it happens in Chiral Perturbation
theory~\cite{chpt}:
the loops arise always at
 $\cO(p^4)$  or higher
($\cO(e^2p^2)$ in the $\gamma\gamma\to w^aw^b$ scattering studied here)
and operators of that chiral dimension are then required
to absorb the UV divergences.
In our case,  $a$ and $v$  are the only relevant $\mL_2$       
parameters for the $\gamma\gamma\to w^aw^b$
scattering amplitude and the related observables studied in this section.

Finally, we would like to mention that one could alternatively extract the running of the   $\mL_4$
chiral parameters
by computing the one-loop UV divergences in the ECLh path integral
by means of the heat-kernel method commonly used
in Chiral Perturbation Theory~\cite{chpt}. However, this ambitious and interesting full computation, is clearly beyond the scope of this work.

\begin{table}[!t]
\begin{center}
\caption{{\small
Set of six observables studied in this article with the ECLh
at   one-loop and their corresponding  relevant
combinations of chiral parameters. The six of them can be  given in terms of the
$\cO(p^2)$ chiral parameter $a$ and three independent  combinations
of $\cO(p^4)$ parameters, $c_\gamma^r$, $a_1^r$ and $(a_2^r-a_3^r)$.
}}
\label{tab.relevant-couplings}
\begin{tabular}{l|cc}
\hline\hline
{\bf Observables}     & \multicolumn{2}{c}{\bf Relevant combinations of parameters} \\
\null  & \ \ from $\mL_2$\ \  & from $\mL_4$ \\
\hline
$\mM(\gamma\gamma\to zz)$     & $a$ & $c_\gamma^{r}$ \\
$\mM(\gamma\gamma\to w^+w^-)$ & $a$ & $(a_1^r- a_2^r+a_3^r),\, c_\gamma^{r}$ \\
$\Gamma(h\to\gamma\gamma)$    & $a$ & $c_\gamma^{r}$ \\
$S$--parameter                & $a$ &
    $a_1^r$
\\
$\mF_{\gamma^*ww}$            & $a$ &
      $(a_2^r-a_3^r)$
\\
$\mF_{\gamma^*\gamma h}$      & --  & $c_\gamma^{r}$\\
\hline\hline
\end{tabular}
\end{center}
\end{table}

\begin{table}[!t]
\begin{center}
\caption{\small
Running of the relevant ECLh parameters and their combinations appearing in the six selected observables. For completeness, we also provide the running of
$a_4^r$ and $a_5^r$ which participate in $ZZ$~and~$W^+W^-$~scattering~\cite{Espriu:2013B,Dobado:2013}.
The third column provides the corresponding running  for the Higgsless ECL case~\cite{Morales:94}.
}
\vspace{.2cm}
\label{tab.running}
\begin{tabular}{ ||c||c|c|| }
\hline \hline
\rule{0pt}{3ex}
 & {\bf ECLh  } &  {\bf ECL}                  
\\
& & (Higgsless)                               
\\[5pt] \hline
\rule{0pt}{3ex}
$\quad \Gamma_{a_1-a_2+a_3}\quad$ & $\quad 0 \quad $ &  0                
\\[5pt] \hline
\rule{0pt}{3ex}
$\quad \Gamma_{c_\gamma}\quad$ & $\quad 0 \quad $ &  -                
\\[5pt] \hline
\rule{0pt}{3ex}
$\quad \Gamma_{a_1}\quad$ & $\quad -\frac{1}{6}(1-a^2) \quad $&$\quad -\frac{1}{6} \quad $                
\\[5pt] \hline
\rule{0pt}{3ex}
$\quad \Gamma_{a_2-a_3} \quad$ & $\quad -\frac{1}{6}(1-a^2) \quad $& $\quad -\, \frac{1}{6} \quad $                  
\\[5pt] \hline
\rule{0pt}{3ex}
$\quad \Gamma_{a_4}\quad$ & $\quad \frac{1}{6}(1-a^2)^2  \quad $ & $\quad \frac{1}{6}\quad $                 
\\[5pt] \hline
\rule{0pt}{3ex}
$\quad \Gamma_{a_5}\quad$ & $\quad \frac{1}{8}(b-a^2)^2
+\frac{1}{12}(1-a^2)^2\quad $ &
$\quad \frac{1}{12}\quad $                 
\\[5pt] \hline
\hline
\end{tabular}
\end{center}
\end{table}

\section{Conclusions}
\label{sec.conclusions}

In this paper we have studied the $\gamma\gamma\to W^+_L W^-_L$ and $\gamma\gamma\to  Z_L Z_L$ scattering processes within the effective chiral Lagrangian approach, including a light Higgs-like scalar as a dynamical field together with the would-be-Goldstone bosons $w^\pm$ and $z$ associated to the electroweak symmetry breaking. We are proposing here the use of
these processes as an optimal tool to discern possible new physics related to the EWSB in the future collider data. We have presented a full one-loop computation of the related amplitudes, by means of the Equivalence Theorem, for the scattering processes $\gamma\gamma\to w^+ w^-$ and $\gamma\gamma\to z z$, which provide a good description of the physical processes of interest here in the kinematic regime ${  m_{W,Z,h} \ll E\ll 4\pi  v  }$.

The computation has been performed up to NLO, which in this chiral Lagrangian context means taking into account all contributing one-loop diagrams generated from ${\cal L}_2$ in addition to the tree level contributions from both
${\cal L}_2$ and ${\cal L}_4$. That means that we have computed for the first time the quantum effects introduced by the light Higgs-like scalar and the would-be-Goldstone bosons $w^\pm$ and $z$ altogether as dynamical fields in the loops of these radiative processes. As part of this computation we have also set clearly here the proper 'chiral counting rules' that are needed to reach a complete NLO result and we have also illustrated the details of the   renormalization procedure involved. For a further check (this, highly non trivial) of our computation we have done the same exercise with two different parametrizations of the $SU(2)_L\times SU(2)_R/SU(2)_{L+R}$ coset, the exponential and the spherical ones, and we have found the same results, as expected.

Our final analytical results, summarized in the equations from Eq.\eqref{LOzz}
through Eq.\eqref{invariant}, are surprisingly very short and extremely simple.
The case of $\gamma \gamma \to zz$ depends just on $a$ and $c_\gamma$,
and these ECLh parameters appear
in the simple form given in Eq.~\eqref{eq.A-zz}. The case of $\gamma \gamma \to w^+w^-$
depends on $a$, $c_\gamma$, $a_1$, $a_2$ and $a_3$, and they also enter
in a very simple way given in Eq.~\eqref{eq.A-ww}. In our opinion,
one of the most relevant features in these simple results,
is the fact that these two amplitudes are found to be given by ECLh parameters
or combinations of them that are renormalization group invariant.
This is a very interesting result and is a consequence of our findings
in the computation of all the one-loop diagrams from the ECLh that when added together
yield a total contribution that is ultraviolet finite,
in both  $\gamma \gamma \to zz$ and
$\gamma \gamma \to w^+w^-$ cases. Specifically, we have found our results in terms
of $a$, $c_\gamma$ and the combination $(a_1-a_2+a_3)$ that do not get
  renormalized,  as it happens in the Higgs-less  ECL case.

It is also worth to remark that the one-loop contributions in our
  final
results show up in the
form  { $(1-a^2)E^2/(16 \pi^2v^2)$.  Since the present fits to LHC data~\cite{LHC-fits}
suggest a value of $a$ close to one,  these corrections are surprisingly suppressed with
respect to the naively expected $E^2/(16 \pi^2 v^2)$ contributions,
typically occurring from chiral loops of chiral effective field theories.
We have tried to understand the origin of this suppression by redoing
the computation in the context
  of the $SO(5)/SO(4)$ MCHM
(appendix~\ref{app.MCHM}), where we have found
  just two contributing one-loop topologies.  Each diagram was suppressed
by $E^2/(16 \pi^2 f^2)$, with $f$ being
 the unique mass-dimension  parameter  of the MCHM $\mL_2$ Lagrangian, being related with
$a$ and $b$ of the ECLh by $v^2/f^2=(1-a^2)=(1-b)/2$.
This comparison with
  the $SO(5)/SO(4)$ MCHM        therefore suggests the existence of a scale
${\Lambda_{\rm ECLh}\sim 4\pi f > 4\pi v }$ which
is the true characteristic cut-off scale of the ECLh.
We also believe that the origin of the simplicity of our results could be relying
on the custodial symmetry invariant structure of the theory and
   an enlarged symmetry
of the dynamical bosons sector ($h$, $w^{\pm},z$) that arises in the relevant Lagrangian for $\gamma\gamma\to w^a w^b$
in the massless Higgs limit.

Finally, regarding the phenomenological relevance of our results, we have selected and studied in this work  a set of four additional related observables: the $h\to\gamma \gamma$ decay width,
the EW precision $S$ parameter, the $\gamma^*\to w^+w^-$ vector form-factor and
the $\gamma^* \gamma\to h$ transition form-factor,
that involve the same subset of chiral parameters as those studied through these work, and whose detailed predictions are collected in Tables 1, 2 and in Appendix~\ref{app.related-obs}. Our proposal for a future phenomenological study is to perform a global analysis of all these four observables together with the two scattering processes explored here, $\gamma\gamma\to w^+ w^-$ and $\gamma\gamma\to z z$. From a future comparison with data, and since these set of six observables  provide an overconstrained system, one could extract
 the values preferred by data for  these involved chiral parameters.
Consequently, this phenomenological analysis could
conclude on the most/least favorable scenarios for the EWSB.

\section*{Acknowledgements}
A. Dobado would like to thank useful conversations with D. Espriu and F.J. Llanes-Estrada.
J.J. Sanz-Cillero thanks A. Pich for useful discussions on the power counting and previous
results both in EW theories and QCD.
This  work is partially supported by the European Union FP7 ITN
INVISIBLES (Marie Curie Actions, PITN- GA-2011- 289442), by the CICYT through the
projects FPA2012-31880, FPA2010-17747,    CSD2007-00042
and FPA2011-27853-C02-01, by the CM (Comunidad Autonoma de Madrid) through the project HEPHACOS S2009/ESP-1473,
by the Spanish Consolider-Ingenio 2010 Programme CPAN (CSD2007-00042)
and by the Spanish MINECO's "Centro de Excelencia Severo Ochoa" Programme under grant SEV-2012-0249.
The work of R.L.~Delgado
is supported by the Spanish MINECO under grant BES-2012-056054.

\appendix

\section{Feynman rules}
\label{app.Feynman}

{  In this Appendix we present the Feynman rules of the ECLh in the two parametrizations,
exponential and spherical. We assume all momenta incoming.   }

\subsection{Vertices from $\mL_2$}

{
\newcommand{\prA}{wiggly}
\newcommand{\ppi}{dashes}
\newcommand{\pphi}{plain}
\newcommand{\bd}[3]{\begin{fmffile}{#1}\begin{fmfgraph}(40,25)\fmfleft{#2}\fmfright{#3}}
\newcommand{\ed}{\end{fmfgraph}\end{fmffile}}
}

\begin{longtable}{c|c|c}
  Vertex
& Exponential   & Spherical   \\\hline\hline\endhead
\parbox[c]{4cm}{\vspace{1cm}\null\hspace{.8cm}%
\begin{fmffile}{fmftempl1}
\begin{fmfgraph*}(50,40)
  \fmfleft{A} \fmflabel{$A^\mu,\,p_1$}{A}
  \fmf{wiggly}{A,v}
  \fmf{dashes}{v,w1}
  \fmf{dashes}{v,w2}
  \fmfright{w2,w1} \fmflabel{$w^+,\,p_2$}{w1} \fmflabel{$w^-,\,p_3$}{w2}
\end{fmfgraph*}
\end{fmffile}\vspace{1cm}}
&
$ie(p_{2\mu}-p_{3\mu})$
&
$ie(p_{2\mu}-p_{3\mu})$
\\\hline
\parbox[c]{4cm}{\vspace{1cm}\null\hspace{.8cm}%
\begin{fmffile}{fmftempl2}
\begin{fmfgraph*}(50,40)
  \fmfleft{R} \fmflabel{$h,\,p_1$}{R}
  \fmf{plain}{R,v}
  \fmf{dashes}{v,w1}
  \fmf{dashes}{v,w2}
  \fmfright{w2,w1} \fmflabel{$w^+,\,p_2$}{w1} \fmflabel{$w^-,\,p_3$}{w2}
\end{fmfgraph*}
\end{fmffile}\vspace{1cm}}
&
$ -\frac{2ia}{v}p_2 p_3 $
&
$ -\frac{2ia}{v}p_2 p_3 $
\\\hline
\parbox[c]{4cm}{\vspace{1cm}\null\hspace{.8cm}%
\begin{fmffile}{fmftempl3}
\begin{fmfgraph*}(50,40)
  \fmfleft{R} \fmflabel{$h,\,p_1$}{R}
  \fmf{plain}{R,v}
  \fmf{dashes}{v,w1}
  \fmf{dashes}{v,w2}
  \fmfright{w2,w1} \fmflabel{$z,\,p_2$}{w1} \fmflabel{$z,\,p_3$}{w2}
\end{fmfgraph*}
\end{fmffile}\vspace{1cm}}
&
$ -\frac{2ia}{v}p_2 p_3 $
&
$ -\frac{2ia}{v}p_2 p_3 $
\\\hline
\parbox[c]{4cm}{\vspace{1cm}\null\hspace{.8cm}%
\begin{fmffile}{fmftempl4}
\begin{fmfgraph*}(50,40)
  \fmfleft{A2,A1} \fmflabel{$A^\mu,\,p_1$}{A1} \fmflabel{$A^\nu,\,p_2$}{A2}
  \fmf{wiggly}{A1,v}
  \fmf{wiggly}{A2,v}
  \fmf{dashes}{v,w1}
  \fmf{dashes}{v,w2}
  \fmfright{w2,w1} \fmflabel{$w^+,\,p_3$}{w1} \fmflabel{$w^-,\,p_4$}{w2}
\end{fmfgraph*}
\end{fmffile}\vspace{1cm}}
&
$ 2ie^2 g_{\mu\nu} $
&
$ 2ie^2 g_{\mu\nu} $
\\\hline
\parbox[c]{4cm}{\vspace{1cm}\null\hspace{.8cm}%
\begin{fmffile}{fmftempl5}
\begin{fmfgraph*}(50,40)
  \fmfleft{R2,A1} \fmflabel{$A^\mu,\,p_1$}{A1} \fmflabel{$h,\,p_2$}{R2}
  \fmf{wiggly}{A1,v}
  \fmf{plain}{R2,v}
  \fmf{dashes}{v,w1}
  \fmf{dashes}{v,w2}
  \fmfright{w2,w1} \fmflabel{$w^+,\,p_3$}{w1} \fmflabel{$w^-,\,p_4$}{w2}
\end{fmfgraph*}
\end{fmffile}\vspace{1cm}}
&
$ \frac{2iae}{v}(p_{3\mu} - p_{4\mu}) $
&
$ \frac{2iae}{v}(p_{3\mu} - p_{4\mu}) $
\\\hline
\parbox[c]{4cm}{\vspace{1cm}\null\hspace{.8cm}%
\begin{fmffile}{fmftempl6}
\begin{fmfgraph*}(50,40)
  \fmfleft{R2,R1} \fmflabel{$h,\,p_1$}{R1} \fmflabel{$h,\,p_2$}{R2}
  \fmf{plain}{R1,v}
  \fmf{plain}{R2,v}
  \fmf{dashes}{v,w1}
  \fmf{dashes}{v,w2}
  \fmfright{w2,w1} \fmflabel{$w^+,\,p_3$}{w1} \fmflabel{$w^-,\,p_4$}{w2}
\end{fmfgraph*}
\end{fmffile}\vspace{1cm}}
&
$ -\frac{2ib}{v^2}p_3 p_4 $
&
$ -\frac{2ib}{v^2}p_3 p_4 $
\\\hline
\parbox[c]{4cm}{\vspace{1cm}\null\hspace{.8cm}%
\begin{fmffile}{fmftempl8}
\begin{fmfgraph*}(50,40)
  \fmfleft{w2,w1} \fmflabel{$w^+,\,p_1$}{w1} \fmflabel{$w^-,\,p_3$}{w2}
  \fmf{dashes}{w1,v}
  \fmf{dashes}{w2,v}
  \fmf{dashes}{v,w3}
  \fmf{dashes}{v,w4}
  \fmfright{w4,w3} \fmflabel{$w^+,\,p_2$}{w3} \fmflabel{$w^-,\,p_4$}{w4}
\end{fmfgraph*}
\end{fmffile}\vspace{1cm}}
&
   $ -\frac{i}{3v^2} [2(p_1p_2 + p_3p_4) + (p_1+p_2)^2] $
&
{    $ -\frac{i}{v^2} [2(p_1p_2 + p_3p_4) - (p_1+p_2)^2] $    }
\\\hline
\parbox[c]{4cm}{\vspace{1cm}\null\hspace{.8cm}%
\begin{fmffile}{fmftempl9}
\begin{fmfgraph*}(50,40)
  \fmfleft{w2,w1} \fmflabel{$z,\,p_2$}{w1} \fmflabel{$z,\,p_3$}{w2}
  \fmf{dashes}{w1,v}
  \fmf{dashes}{w2,v}
  \fmf{dashes}{v,w3}
  \fmf{dashes}{v,w4}
  \fmfright{w4,w3} \fmflabel{$w^+,\,p_1$}{w3} \fmflabel{$w^-,\,p_4$}{w4}
\end{fmfgraph*}
\end{fmffile}\vspace{1cm}}
&
$ \frac{i}{3v^2} [2(p_1p_4 + p_2p_3) + (p_1 + p_4)^2] $ 
&
$ \frac{i}{v^2} (p_1 + p_4)^2 $
\\\hline
\parbox[c]{4cm}{\vspace{1cm}\null\hspace{.8cm}%
\begin{fmffile}{fmftempl10}
\begin{fmfgraph*}(50,40)
  \fmfleft{w2,w1} \fmflabel{$z,\,p_1$}{w1} \fmflabel{$z,\,p_2$}{w2}
  \fmf{dashes}{w1,v}
  \fmf{dashes}{w2,v}
  \fmf{dashes}{v,w3}
  \fmf{dashes}{v,w4}
  \fmfright{w4,w3} \fmflabel{$z,\,p_3$}{w3} \fmflabel{$z,\,p_4$}{w4}
\end{fmfgraph*}
\end{fmffile}\vspace{1cm}}
&
$0$
&
$ -\frac{2i}{v^2} [p_1p_4 + p_2p_3 - (p_1+p_4)^2] $
\\\hline
\parbox[c]{4cm}{\vspace{1cm}\null\hspace{.8cm}%
\begin{fmffile}{fmftempl11}
\begin{fmfgraph*}(50,40)
  \fmfleft{R2,R1} \fmflabel{$h,\,p_1$}{R1} \fmflabel{$h,\,p_2$}{R2}
  \fmf{plain}{R1,v}
  \fmf{plain}{R2,v}
  \fmf{dashes}{v,w3}
  \fmf{dashes}{v,w4}
  \fmfright{w4,w3} \fmflabel{$z,\,p_3$}{w3} \fmflabel{$z,\,p_4$}{w4}
\end{fmfgraph*}
\end{fmffile}\vspace{1cm}}
&
$-\frac{2ib}{v^2}p_3p_4$
&
$-\frac{2ib}{v^2}p_3p_4$
\\\hline
\parbox[c]{4cm}{\vspace{1cm}\null\hspace{.8cm}%
\begin{fmffile}{fmftempl12}
\begin{fmfgraph*}(50,40)
  \fmfleft{w2,w1} \fmflabel{$w^+,\,p_4$}{w1} \fmflabel{$w^-,\,p_5$}{w2}
  \fmf{dashes}{w1,v}
  \fmf{dashes}{w2,v}
  \fmf{wiggly}{v,A2}
  \fmf{wiggly}{v,A3}
  \fmf{plain}{v,R4}
  \fmfright{R4,A3,A2} \fmflabel{$A^\mu,\,p_1$}{A2} \fmflabel{$A^\nu,\,p_2$}{A3} \fmflabel{$h,\,p_3$}{R4}
\end{fmfgraph*}
\end{fmffile}\vspace{1cm}}
&
{    $ \frac{4iae^2}{v}g_{\mu\nu}$    }
&
{    $\frac{4iae^2}{v}g_{\mu\nu}$     }
\\\hline
\parbox[c]{4cm}{\vspace{1cm}\null\hspace{.8cm}%
\begin{fmffile}{fmftempl13}
\begin{fmfgraph*}(50,40)
  \fmfleft{w2,w1} \fmflabel{$w^+,\,p_4$}{w1} \fmflabel{$w^-,\,p_5$}{w2}
  \fmf{dashes}{w1,v}
  \fmf{dashes}{w2,v}
  \fmf{wiggly}{v,A2}
  \fmf{plain}{v,R3}
  \fmf{plain}{v,R4}
  \fmfright{R4,R3,A2} \fmflabel{$A^\mu,\,p_1$}{A2} \fmflabel{$h,\,p_2$}{R3} \fmflabel{$h,\,p_3$}{R4}
\end{fmfgraph*}
\end{fmffile}\vspace{1cm}}
&
$\frac{2ibe}{v^2}(p_{4\mu}-p_{5\mu})$
&
$\frac{2ibe}{v^2}(p_{4\mu}-p_{5\mu})$
\\\hline
\parbox[c]{4cm}{\vspace{1cm}\null\hspace{.8cm}%
\begin{fmffile}{fmftempl14}
\begin{fmfgraph*}(50,40)
  \fmfleft{w2,w1} \fmflabel{$w^+,\,p_2$}{w1} \fmflabel{$w^-,\,p_5$}{w2}
  \fmf{dashes}{w1,v}
  \fmf{dashes}{w2,v}
  \fmf{wiggly}{v,A}
  \fmf{dashes}{v,W3}
  \fmf{dashes}{v,W4}
  \fmfright{W4,W3,A} \fmflabel{$A^\mu,\,p_1$}{A} \fmflabel{$z,\,p_3$}{W3} \fmflabel{$z,\,p_4$}{W4}
\end{fmfgraph*}
\end{fmffile}\vspace{1cm}}
&
$\frac{2ie}{3v^2}(p_{5\mu}-p_{2\mu})$
&
$0$
\\\hline
\parbox[c]{4cm}{\vspace{1cm}\null\hspace{.8cm}%
\begin{fmffile}{fmftempl15}
\begin{fmfgraph*}(50,40)
  \fmfleft{w2,w1} \fmflabel{$w^+,\,p_2$}{w1} \fmflabel{$w^-,\,p_5$}{w2}
  \fmf{dashes}{w1,v}
  \fmf{dashes}{w2,v}
  \fmf{wiggly}{v,A}
  \fmf{dashes}{v,W3}
  \fmf{dashes}{v,W4}
  \fmfright{W4,W3,A} \fmflabel{$A^\mu,\,p_1$}{A} \fmflabel{$w^+,\,p_3$}{W3} \fmflabel{$w^-,\,p_4$}{W4}
\end{fmfgraph*}
\end{fmffile}\vspace{1cm}}
&
$\frac{4ie}{3v^2}(p_{5\mu} + p_{4\mu} - p_{3\mu} - p_{2\mu})$
&
$0$
\\\hline
\parbox[c]{4cm}{\vspace{1cm}\null\hspace{.8cm}%
\begin{fmffile}{fmftempl16}
\begin{fmfgraph*}(50,40)
  \fmfleft{R1,A2,A1} \fmflabel{$A^\mu,\,p_1$}{A1} \fmflabel{$A^\nu,\,p_2$}{A2} \fmflabel{$h,\,p_3$}{R1}
  \fmf{plain}{R1,v}
  \fmf{plain}{v,R2}
  \fmf{wiggly}{A1,v}
  \fmf{wiggly}{A2,v}
  \fmf{dashes}{v,W1}
  \fmf{dashes}{v,W2}
  \fmfright{R2,W2,W1} \fmflabel{$w^+,\,p_5$}{W1} \fmflabel{$w^-,\,p_6$}{W2} \fmflabel{$h,\,p_4$}{R2}
\end{fmfgraph*}
\end{fmffile}\vspace{1cm}}
&
$\frac{4ibe^2}{v^2}g_{\mu\nu}$
&
$\frac{4ibe^2}{v^2}g_{\mu\nu}$
\\\hline
\end{longtable}

\subsection{Vertices from $\mL_4$}

\begin{longtable}{c|c|c}
Vertex
& Exponential   & Spherical   \\\hline\hline\endhead
\parbox[c]{4cm}{\vspace{1cm}\null\hspace{.8cm}%
\begin{fmffile}{fmftempl17}
\begin{fmfgraph*}(50,40)
  \fmfleft{A} \fmflabel{$A^\mu,\,p_1$}{A}
  \fmf{wiggly}{A,v}
  \fmf{dashes}{v,w1}
  \fmf{dashes}{v,w2}
  \fmfright{w2,w1} \fmflabel{$w^+,\,p_2$}{w1} \fmflabel{$w^-,\,p_3$}{w2}
  \fmfv{decor.shape=square,decor.filled=30,decor.size=4thick}{v}
\end{fmfgraph*}
\end{fmffile}\vspace{1cm}}
&
{    $ - \frac{4 i e (a_3-a_2)}{v^2} \bigg( (p_1 p_3) p_{2\, \mu} -(p_1 p_2) p_{3\, \mu}\bigg)$   }
&
{    $ -  \frac{4 i e (a_3-a_2)}{v^2} \bigg( (p_1 p_3) p_{2\, \mu} -(p_1 p_2) p_{3\, \mu}\bigg)$   }
\\\hline
\parbox[c]{4cm}{\vspace{1cm}\null\hspace{.8cm}%
\begin{fmffile}{fmftempl18}
\begin{fmfgraph*}(50,40)
  \fmfleft{A2,A1} \fmflabel{$A^\mu,\,p_1$}{A1} \fmflabel{$A^\nu,\,p_2$}{A2}
  \fmf{wiggly}{A1,v}
  \fmf{wiggly}{A2,v}
  \fmf{dashes}{v,w1}
  \fmf{dashes}{v,w2}
  \fmfright{w2,w1} \fmflabel{$w^+,\,p_3$}{w1} \fmflabel{$w^-,\,p_4$}{w2}
  \fmfv{decor.shape=square,decor.filled=30,decor.size=4thick}{v}
\end{fmfgraph*}
\end{fmffile}\vspace{1cm}}
&
$\begin{array}{l}
\frac{8ie^2 a_1}{v^2} \bigg((p_1 p_2) g_{\mu\nu} - p_{2\, \mu} p_{1\, \nu}\bigg)
\\ \\
+ \frac{4ie^2 (a_3-a_2)}{v^2} \bigg((p_1+ p_2)^2 g_{\mu\nu}
\\ \qquad\quad - (p_{1\, \mu} + p_{2\, \mu}) p_{1\, \nu}
\\ \qquad\quad  - p_{2\, \mu} (p_{1\, \nu}+p_{2\, \nu})
\bigg)
\end{array}   $
&
$\begin{array}{l}
\frac{8ie^2 a_1}{v^2} \bigg((p_1 p_2) g_{\mu\nu} - p_{2\, \mu} p_{1\, \nu}\bigg)
\\ \\
+ \frac{4ie^2 (a_3-a_2)}{v^2} \bigg((p_1+ p_2)^2 g_{\mu\nu}
\\ \qquad\quad - (p_{1\, \mu} + p_{2\, \mu}) p_{1\, \nu}
\\ \qquad\quad - p_{2\, \mu} (p_{1\, \nu}+p_{2\, \nu})
\bigg)
\end{array}   $
\\\hline
\parbox[c]{4cm}{\vspace{1cm}\null\hspace{.8cm}%
\begin{fmffile}{fmftempl-cgamB}
\begin{fmfgraph*}(50,40)
  \fmfleft{A2,A1} \fmflabel{$A^\mu,\,p_1$}{A1} \fmflabel{$A^\nu,\,p_2$}{A2}
  \fmf{wiggly}{A1,v}
  \fmf{wiggly}{A2,v}
  \fmf{plain}{v,h}
  \fmfright{h} \fmflabel{$h,\,p_3$}{h}
  \fmfv{decor.shape=square,decor.filled=30,decor.size=4thick}{v}
\end{fmfgraph*}
\end{fmffile}\vspace{1cm}}
&
$ \frac{2i c_\gamma}{v} \left( (p_1p_2) g_{\mu\nu} - p_{2\, \mu} p_{1\, \nu}\right)$
&
$ \frac{2i c_\gamma}{v} \left( (p_1p_2) g_{\mu\nu} - p_{2\, \mu} p_{1\, \nu}\right)$
\\\hline
\end{longtable}

\section{Contribution from each diagram  to the $\gamma\gamma\to w^aw^b$ amplitudes}
\label{app.diagrams}

\subsection{$\gamma\gamma\to zz$ scattering amplitude}

In both the exponential and spherical parametrizations
the non-vanishing diagrams in our one-loop $\gamma(k_1,\epsilon_1)\gamma(k_2,\epsilon_2)\to z(p_1)z(p_2)$ computation yield
\begin{eqnarray}
\mM^{\rm 1} &=&  -\frac{i e^2 (s B_0\text{(s,0,0)} \left(\epsilon _1\epsilon _2\right)+s \left(\epsilon _1\epsilon _2\right)-2 \left(\epsilon _1k_2\right) \left(\epsilon _2k_1\right))}{16 \pi ^2 v^2}
\, ,\\
\mM^{\rm 2} &=&-\frac{i e^2 (s B_0\text{(s,0,0)} \left(\epsilon _1\epsilon _2\right)+s \left(\epsilon _1\epsilon _2\right)-2 \left(\epsilon _1k_2\right) \left(\epsilon _2k_1\right))}{16 \pi ^2 v^2}
\, ,\\
\mM^{\rm 3} &=&-\frac{i a^2 e^2 (B_0\text{(s,0,0)} \left(\epsilon _1\epsilon _2\right) (t+u)+2 \left(\epsilon _1k_2\right) \left(\epsilon _2k_1\right)+\left(\epsilon _1\epsilon _2\right) (t+u))}{16 \pi ^2 v^2}
\, ,\\
\mM^{\rm 4} &=&-\frac{i a^2 e^2 (B_0\text{(s,0,0)} \left(\epsilon _1\epsilon _2\right) (t+u)+2 \left(\epsilon _1k_2\right) \left(\epsilon _2k_1\right)+\left(\epsilon _1\epsilon _2\right) (t+u))}{16 \pi ^2 v^2}
\, ,\\
\mM^{\rm 6} &=&\frac{i e^2 s B_0\text{(s,0,0)} \left(\epsilon _1\epsilon _2\right)}{8 \pi ^2 v^2}
\, ,\\
\mM^{\rm 10} &=&-\frac{i a^2 e^2 s B_0\text{(s,0,0)} \left(\epsilon _1\epsilon _2\right)}{8 \pi ^2 v^2}
\, ,
\end{eqnarray}
  with the Mandelstam variables defined as usual, $s=(p_1+p_2)^2$, $t=(k_1-p_1)^2$
and $u=(k_1-p_2)^2$, the relevant momentum combination is defined as $\Delta^\mu\equiv p_1^\mu -p_2^\mu$,
and the $\epsilon_i$'s are the  polarization vectors of the external photons.
 The vanishing of the diagrams 5, 7, 8 and 9 is implied by the fact  that
we work in the Landau gauge, the Higgs mass is taken to be zero
and the incoming photons are set on-shell.
Furthermore, in the spherical parametrization, the diagrams 5 and 7 are always
absent as there is no $\gamma\omega\omega\omega\omega$ vertex in these coordinates.
In order to reach our final expression for the total amplitude we used the
on-shell kinematical condition $s+t+u=0$.   }
For the relevant massless  Feynman integral here we follow the notation
\bear
B_0(q^2,0,0)&=&\Int\Frac{d^dk}{i\pi^2} \Frac{1}{k^2 \, (q-k)^2} \, .
\eear

We would like also to notice that the total result of the one-loop contributions to the $\gamma\gamma\to zz$
is in agreement with the recent result in Ref.~\cite{Talavera:2014} within the QCD context for $\gamma\gamma\to\pi^0\pi^0$
including both the pions and a light scalar singlet $S_1$ when their masses $m_\pi$ and $m_{S_1}$ are set to zero
and $c_\gamma=0$.

\subsection{$\gamma\gamma\to w^+w^-$ scattering amplitude}

In both the exponential and spherical parametrizations
the non-vanishing diagrams in our one-loop
$\gamma(k_1,\epsilon_1)\gamma(k_2,\epsilon_2)\to w^+(p_1) w^-(p_2)$
computation yield
\begin{eqnarray}
\mM^{\rm 1} &=&-\frac{i e^2}{144 \pi ^2 s v^2}
\bigg(3 B_0\text{(s,0,0)} (t+u) (-\left(\epsilon _1\text{$\Delta $)}\right.
\left(\epsilon _2k_1\right)+\left(\epsilon _1k_2\right)
\left(\epsilon _2\text{$\Delta $)}\right.+\left(\epsilon _1\epsilon _2\right) t
\\
&&   \qquad\quad +2 \left(\epsilon _1\epsilon _2\right) u)
+2 \left(\epsilon _1k_2\right) (\left(\epsilon _2\text{$\Delta $)}\right. (t+u)
\nn\\
&&   \qquad\quad+3 \left(\epsilon _2k_1\right) (t+2 u))+(t+u) (-2 \left(\epsilon _1\text{$\Delta $)}\right. \left(\epsilon _2k_1\right)+2 \left(\epsilon _1\epsilon _2\right) t
+7 \left(\epsilon _1\epsilon _2\right) u)\bigg)
\, ,\nn\\
\mM^{\rm 2} &=&-\frac{i e^2}{144 \pi ^2 s v^2}
\bigg(3 B_0\text{(s,0,0)} (t+u) (\left(\epsilon _1\text{$\Delta $)}\right.
\left(\epsilon _2k_1\right)-\left(\epsilon _1k_2\right)
\left(\epsilon _2\text{$\Delta $)}\right.+2 \left(\epsilon _1\epsilon _2\right) t
+\left(\epsilon _1\epsilon _2\right) u)
\nn\\
&&   \qquad\quad -2 \left(\epsilon _1k_2\right) (\left(\epsilon _2\text{$\Delta $)}\right. (t+u)-3 \left(\epsilon _2k_1\right) (2 t+u))+(t+u) (2 \left(\epsilon _1\text{$\Delta $)}\right. \left(\epsilon _2k_1\right)
\nn\\
&&   \qquad\quad +7 \left(\epsilon _1\epsilon _2\right) t+2 \left(\epsilon _1\epsilon _2\right) u)\bigg)
\, ,\\
\mM^{\rm 3} &=& \frac{i a^2 e^2 }{288 \pi ^2 v^2}
\bigg(3 B_0\text{(t,0,0)} (2 \left(\epsilon _1\epsilon _2\right) t
-5 (\left(\epsilon _1\text{$\Delta $)}\right.+\left(\epsilon _1k_2\right))
(\left(\epsilon _2\text{$\Delta $)}\right.-\left(\epsilon _2k_1\right)))
\nn\\
&&\qquad\quad +(-\left(\epsilon _1\text{$\Delta $)}\right.-\left(\epsilon _1k_2\right)) (\left(\epsilon _2\text{$\Delta $)}\right.-\left(\epsilon _2k_1\right))+4 \left(\epsilon _1\epsilon _2\right) t\bigg)
\, ,\\
\mM^{\rm 4} &=&\frac{i a^2 e^2}{288 \pi ^2 v^2}
\bigg(3 B_0\text{(u,0,0)} (2 \left(\epsilon _1\epsilon _2\right) u
-5 (\left(\epsilon _1\text{$\Delta $)}\right.-\left(\epsilon _1k_2\right))
 (\left(\epsilon _2\text{$\Delta $)}\right.+\left(\epsilon _2k_1\right)))
\nn\\
&&\qquad\quad   -\left(\epsilon _1\text{$\Delta $)}\right. (\left(\epsilon _2\text{$\Delta $)}\right.+\left(\epsilon _2k_1\right))+\left(\epsilon _1k_2\right) (\left(\epsilon _2\text{$\Delta $)}\right.+\left(\epsilon _2k_1\right))+4 \left(\epsilon _1\epsilon _2\right) u\bigg)
\, ,\\
\mM^{\rm 5} &=&\frac{i a^2 e^2}{288 \pi ^2 v^2}
\bigg(3 B_0\text{(u,0,0)} (2 \left(\epsilon _1\epsilon _2\right) u
-5 (\left(\epsilon _1\text{$\Delta $)}\right.-\left(\epsilon _1k_2\right))
(\left(\epsilon _2\text{$\Delta $)}\right.+\left(\epsilon _2k_1\right)))
\nn\\
&&\qquad\quad  -\left(\epsilon _1\text{$\Delta $)}\right. (\left(\epsilon _2\text{$\Delta $)}\right.+\left(\epsilon _2k_1\right))+\left(\epsilon _1k_2\right) (\left(\epsilon _2\text{$\Delta $)}\right.+\left(\epsilon _2k_1\right))+4 \left(\epsilon _1\epsilon _2\right) u\bigg)
\, ,
\end{eqnarray}
\begin{eqnarray}
\mM^{\rm 6} &=&\frac{i a^2 e^2}{288 \pi ^2 v^2}
\bigg(3 B_0\text{(t,0,0)} (2 \left(\epsilon _1\epsilon _2\right) t
-5 (\left(\epsilon _1\text{$\Delta $)}\right.+\left(\epsilon _1k_2\right))
(\left(\epsilon _2\text{$\Delta $)}\right.-\left(\epsilon _2k_1\right)))
\nn\\
&&\qquad\quad  +(-\left(\epsilon _1\text{$\Delta $)}\right.-\left(\epsilon _1k_2\right))
 (\left(\epsilon _2\text{$\Delta $)}\right.-\left(\epsilon _2k_1\right))
 +4 \left(\epsilon _1\epsilon _2\right) t\bigg)
\, ,\\
\mM^{\rm 7} &=&\frac{i a^2 e^2 s B_0\text{(s,0,0)} \left(\epsilon _1\epsilon _2\right)}{16 \pi ^2 v^2}
\, ,\\
\mM^{\rm 8} &=&\frac{i a^2 e^2 B_0\text{(t,0,0)} (\left(\epsilon _1\text{$\Delta $)}\right.+\left(\epsilon _1k_2\right)) (\left(\epsilon _2\text{$\Delta $)}\right.-\left(\epsilon _2k_1\right))}{32 \pi ^2 v^2}
\, ,\\
\mM^{\rm 9} &=&\frac{i a^2 e^2 B_0\text{(u,0,0)} (\left(\epsilon _1\text{$\Delta $)}\right.-\left(\epsilon _1k_2\right)) (\left(\epsilon _2\text{$\Delta $)}\right.+\left(\epsilon _2k_1\right))}{32 \pi ^2 v^2}
\, ,\\
\mM^{\rm 10} &=&\frac{i a^2 e^2 B_0\text{(u,0,0)} (\left(\epsilon _1\text{$\Delta $)}\right.-\left(\epsilon _1k_2\right)) (\left(\epsilon _2\text{$\Delta $)}\right.+\left(\epsilon _2k_1\right))}{32 \pi ^2 v^2}
\, ,
\end{eqnarray}
\begin{eqnarray}
\mM^{\rm 11} &=&\frac{i a^2 e^2 B_0\text{(t,0,0)} (\left(\epsilon _1\text{$\Delta $)}\right.+\left(\epsilon _1k_2\right)) (\left(\epsilon _2\text{$\Delta $)}\right.-\left(\epsilon _2k_1\right))}{32 \pi ^2 v^2}
\, ,\\
\mM^{\rm 12} &=&-\frac{i a^2 e^2 (B_0\text{(s,0,0)} \left(\epsilon _1\epsilon _2\right) (t+u)+2 \left(\epsilon _1k_2\right) \left(\epsilon _2k_1\right)+\left(\epsilon _1\epsilon _2\right) (t+u))}{16 \pi ^2 v^2}
\, ,\\
\mM^{\rm 13} &=&-\frac{i a^2 e^2 (B_0\text{(s,0,0)} \left(\epsilon _1\epsilon _2\right) (t+u)+2 \left(\epsilon _1k_2\right) \left(\epsilon _2k_1\right)+\left(\epsilon _1\epsilon _2\right) (t+u))}{16 \pi ^2 v^2}
\, ,\\
\mM^{\rm 14} &=&\frac{i a^2 e^2 (t+u)}{288 \pi ^2 s^2 v^2}
\bigg(6 (t+u) (B_0\text{(s,0,0)} (\left(\epsilon _1\text{$\Delta $)}\right.
\left(\epsilon _2k_1\right)-\left(\epsilon _1k_2\right)
 \left(\epsilon _2\text{$\Delta $)}\right.+2 \left(\epsilon _1\epsilon _2\right) t
\nn\\
&&\qquad\quad
 +\left(\epsilon _1\epsilon _2\right) u)+B_0\text{(t,0,0)}
  ((\left(\epsilon _1\text{$\Delta $)}\right.+\left(\epsilon _1k_2\right))
  (\left(\epsilon _2\text{$\Delta $)}\right.-\left(\epsilon _2k_1\right))
  -\left(\epsilon _1\epsilon _2\right) t))
\nn\\
&&\qquad\quad
 +\left(\epsilon _1\text{$\Delta $)}\right.
  (\left(\epsilon _2\text{$\Delta $)}\right.+3 \left(\epsilon _2k_1\right)) (t+u)
  +\left(\epsilon _1k_2\right) (\left(\epsilon _2k_1\right) (23 t+11 u)
\nn\\
&&\qquad\quad
  -3 \left(\epsilon _2\text{$\Delta $)}\right. (t+u))
  +2 \left(\epsilon _1\epsilon _2\right) (5 t+2 u) (t+u)\bigg)
\, ,\\
\mM^{\rm 15} &=&\frac{i a^2 e^2 (t+u)}{288 \pi ^2 s^2 v^2}
\bigg(6 (t+u) (B_0\text{(s,0,0)} (-\left(\epsilon _1\text{$\Delta $)}\right.
\left(\epsilon _2k_1\right)+\left(\epsilon _1k_2\right)
 \left(\epsilon _2\text{$\Delta $)}\right.+\left(\epsilon _1\epsilon _2\right) t
\nn\\
&&\qquad\quad  +2 \left(\epsilon _1\epsilon _2\right) u)+B_0\text{(u,0,0)}
 ((\left(\epsilon _1\text{$\Delta $)}\right.-\left(\epsilon _1k_2\right))
  (\left(\epsilon _2\text{$\Delta $)}\right.+\left(\epsilon _2k_1\right))
  -\left(\epsilon _1\epsilon _2\right) u))
\nn\\
&&\qquad\quad  +\left(\epsilon _1\text{$\Delta $)}\right.
(\left(\epsilon _2\text{$\Delta $)}\right.
  -3 \left(\epsilon _2k_1\right)) (t+u)+\left(\epsilon _1k_2\right)
   (3 \left(\epsilon _2\text{$\Delta $)}\right. (t+u)
\nn\\
&&\qquad\quad   +\left(\epsilon _2k_1\right) (11 t+23 u))
   +2 \left(\epsilon _1\epsilon _2\right) (2 t+5 u) (t+u)\bigg)
\, ,
\end{eqnarray}
\begin{eqnarray}
\mM^{\rm 16} &=& \frac{i e^2 s B_0\text{(s,0,0)}
\left(\epsilon _1\epsilon _2\right)}{16 \pi ^2 v^2}
\, ,\\
\mM^{\rm 17} &=&\frac{i a^2 e^2}{288 \pi ^2 v^2}
\bigg(6 B_0\text{(t,0,0)} (7 (\left(\epsilon _1\text{$\Delta $)}\right.
+\left(\epsilon _1k_2\right)) (\left(\epsilon _2\text{$\Delta $)}\right.
-\left(\epsilon _2k_1\right))-\left(\epsilon _1\epsilon _2\right) t)
\nn\\
&&\qquad\quad    +(\left(\epsilon _1\text{$\Delta $)}\right.+\left(\epsilon _1k_2\right)) (\left(\epsilon _2\text{$\Delta $)}\right.-\left(\epsilon _2k_1\right))-4 \left(\epsilon _1\epsilon _2\right) t\bigg)
\, ,\\
\mM^{\rm 18} &=&\frac{i a^2 e^2}{288 \pi ^2 v^2}
\bigg(6 B_0\text{(u,0,0)} (7 (\left(\epsilon _1\text{$\Delta $)}\right.
-\left(\epsilon _1k_2\right)) (\left(\epsilon _2\text{$\Delta $)}\right.
+\left(\epsilon _2k_1\right))-\left(\epsilon _1\epsilon _2\right) u)
\nn\\
&&\qquad\quad +(\left(\epsilon _1\text{$\Delta $)}\right.-\left(\epsilon _1k_2\right)) (\left(\epsilon _2\text{$\Delta $)}\right.+\left(\epsilon _2k_1\right))-4 \left(\epsilon _1\epsilon _2\right) u\bigg)
\, ,\\
\mM^{\rm 19} &=&-\frac{3 i a^2 e^2 B_0\text{(t,0,0)} (\left(\epsilon _1\text{$\Delta $)}\right.+\left(\epsilon _1k_2\right)) (\left(\epsilon _2\text{$\Delta $)}\right.-\left(\epsilon _2k_1\right))}{32 \pi ^2 v^2}
\, ,\\
\mM^{\rm 20} &=&-\frac{3 i a^2 e^2 B_0\text{(u,0,0)} (\left(\epsilon _1\text{$\Delta $)}\right.-\left(\epsilon _1k_2\right)) (\left(\epsilon _2\text{$\Delta $)}\right.+\left(\epsilon _2k_1\right))}{32 \pi ^2 v^2}
\, ,
\end{eqnarray}
\begin{eqnarray}
\mM^{\rm 21} &=&-\frac{3 i a^2 e^2 B_0\text{(u,0,0)} (\left(\epsilon _1\text{$\Delta $)}\right.-\left(\epsilon _1k_2\right)) (\left(\epsilon _2\text{$\Delta $)}\right.+\left(\epsilon _2k_1\right))}{32 \pi ^2 v^2}
\, ,\\
\mM^{\rm 22} &=&-\frac{3 i a^2 e^2 B_0\text{(t,0,0)} (\left(\epsilon _1\text{$\Delta $)}\right.+\left(\epsilon _1k_2\right)) (\left(\epsilon _2\text{$\Delta $)}\right.-\left(\epsilon _2k_1\right))}{32 \pi ^2 v^2}
\, ,\\
\mM^{\rm 23} &=&-\frac{i a^2 e^2 s B_0\text{(s,0,0)} \left(\epsilon _1\epsilon _2\right)}{8 \pi ^2 v^2}
\, ,\\
\mM^{\rm 24} &=&\frac{i a^2 e^2 B_0\text{(t,0,0)} (\left(\epsilon _1\text{$\Delta $)}\right.+\left(\epsilon _1k_2\right)) (\left(\epsilon _2\text{$\Delta $)}\right.-\left(\epsilon _2k_1\right))}{16 \pi ^2 v^2}
\, ,\\
\mM^{\rm 25} &=&\frac{i a^2 e^2 B_0\text{(u,0,0)} (\left(\epsilon _1\text{$\Delta $)}\right.-\left(\epsilon _1k_2\right)) (\left(\epsilon _2\text{$\Delta $)}\right.+\left(\epsilon _2k_1\right))}{16 \pi ^2 v^2}
\, ,
\end{eqnarray}

 with   $s$, $t$, $u$, $\Delta$  and the $\epsilon_i$'s
defined as in the previous section.
The remaining diagrams are zero in both coset coordinates when we work
in the Landau gauge, the Higgs mass is taken to be zero
and the incoming photons are set on-shell.
Furthermore, in the spherical parametrization, the diagrams 26 and  27  are always
absent as there is no $\gamma\omega\omega\omega\omega$ vertex in these coordinates.
 In order to reach our final expression for the total amplitude we used the
on-shell kinematical condition $s+t+u=0$.

\section{One-loop $\gamma\gamma  \to zz$
and $\gamma \gamma \to w^+ w^-$  scattering in           
MCHM}
\label{app.MCHM}

This appendix is devoted to the computation of the
  one-loop
amplitudes considered in this work for the $\gamma \gamma \to z z$ and  $\gamma \gamma \to w^+ w^-$
processes in the context of the so called
   $SO(5)/SO(4)$ MCHM~\cite{MCHM}.
In this model it is assumed that some global symmetry breaking takes place at some scale
  $4 \pi f>  4\pi v$
so that the group $G=SO(5)$ is spontaneously broken to the subgroup $H=SO(4)$. Therefore  the corresponding Goldstone bosons (GBs)
live in the coset $K=G/H=S^4$. These four GBs will be identified with the Higgs-like boson $h$ and the three WBGBs needed for giving masses to the $W^\pm$ and $Z$. The
$G$ group contains also the subgroup $H'=SO(4)=SU(2)_L\times SU(2)_R$
in such a way that the gauge group  $H_g=SU(2)_L \times U(1)_Y$ is a subgroup of $H'$. Notice however that $H \ne H'$. In fact $G=SO(5)$ has many $SO(4)$
subgroups which can be defined by giving a fixed five dimensional vector belonging
to the $G$ fundamental representation which is invariant under the action of the $G$  subgroup.
For example the $H'$ group is defined by
 the invariant vector $\Phi_0'$ and similarly $H$ is defined by $\Phi_0$:
\bear
\Phi_0' = f \left(\begin{array}{c}0 \\ 0\\ 0\\ 0 \\ 1\end{array}\right)\, ,
\qquad \qquad
\Phi_0 =  f \left(\begin{array}{c}0 \\ 0\\ 0\\ s \\ c \end{array}\right)\, ,
\eear
with $s= \sin \theta$,   $c=\cos \theta $ and  $\theta $ being the misalignment angle. Thus the $H'$ group acts only on the first four components of any five dimensional $\Phi$ vector  belonging to the $G$ fundamental representation.
The $SU(2)_L$ subgroup has generators $T^k_L=i M^k_L/2$ ($k=1,2,3$):
\bear
&    M_L^1 = \left(\begin{array}{ccccc}
0 & 0&0&-&0 \\
0 &0&-&0&0 \\
0&+&0 & 0&0 \\
+&0 & 0 &0&0 \\
0&0&0&0&0
\end{array}\right)\, ,
\qquad \qquad
M_L^2 = \left(\begin{array}{ccccc}
0 & 0 &+&0&0 \\
0 &0&0&-&0 \\
-&0&0 & 0 &0 \\
0&+ & 0 &0&0 \\
0&0&0&0&0
\end{array}\right)\, ,
& \nn\\  &
 M_L^3   =  \left(\begin{array}{ccccc}
0 & -&0&0&0 \\
+    &0&0&0&0 \\
0&0&0 & -  &0 \\
0&0 &+ &0&0 \\
0&0&0&0&0
\end{array}\right)\, ,    &
\eear
and the $U(1)_Y$ is generated by the third $SU(2)_R$ generator which is given by
$T_R^3= i M_R^3/2=  i M_Y/2$ with:
   \bear
M_Y &=& \left(\begin{array}{ccccc}
0 & -&0&0&0 \\
+ &0&0&0&0 \\
0&0&0 & + &0 \\
0&0 & -&0&0 \\
0&0&0&0&0
\end{array}\right)\, .
\eear
Clearly these generators fulfill $[T_L^i,T_L^j]= i \epsilon _{ijk}T_L^k$ and $[T_L^k,T_Y]=0$.
 The expressions for all the $SO(5)$ generators can be found in the appendices
of Refs.~\cite{MCHM}.

Now the low energy dynamics of the system can be described
by the non-linear sigma model (NL$\sigma$M)
given by the Lagrangian:
\bear
  \mL^{\rm MCHM}_2
&=& \Frac{1}{2} \partial^\mu \Phi^{\,\, \dagger} \, \partial_\mu\Phi\, \mid_{S^4} \, ,
\label{eq.Phi-vector}
\eear
with the $G$ fundamental representation vector parametrized as:
\bear
  \Phi
&=& \left(\begin{array}{c}\omega^1 \\ \omega^2\\ \omega^3\\ c \omega^4+s \chi \\  -s\omega^4+ c\chi \end{array}\right)\, .
\eear
The condition for $\Phi$ being  in $S^4$ is just $\Phi^T \Phi  = f^2$ from which we can obtain the fifth coordinate $\chi$ as a function of the first four $\omega^\alpha$  ($\alpha=1,2,3,4$):
\bear
\chi= \left(f^2 -\sum_\alpha (\omega^\alpha)^2\right)^{1/2}.
\eear
Therefore we have the $G/H=SO(5)/SO(4)=S^4$ NL$\sigma$M Lagrangian:
\bear
    \mL^{\rm MCHM}_2(\omega^\alpha)
 &=& \Frac{1}{2} g_{\alpha\beta}\partial^\mu \omega^\alpha\, \partial_\mu \omega^\beta
\eear
with the $S^4$ metric being given in our coordinates by:
\bear
 g_{\alpha\beta}  & =  & \delta_{\alpha\beta}+\frac{\omega^\alpha\omega^\beta}{f^2-\sum_\alpha (\omega^\alpha)^2}.
\eear

Now we can introduce $SU(2)_L\times U(1)_Y$ gauge interactions by introducing the covariant
derivative:
\bear
D_\mu & = & \partial_\mu - i g T_L^k W_\mu^k
-i g'T_Y B_\mu
\eear
where $W_\mu^k$ and $B_\mu$ are the $SU(2)_L$ and $U(1)_Y$ gauge bosons respectively.
 The photon field is then given as usual by
\bear
A_\mu & = & \sin\theta_W \, W^3_\mu \, +\, \cos\theta_W\, B_\mu\, ,
\eear
with $\theta_W$ being the Weinberg angle: $\sin\theta_W = g' / \sqrt{g^2+g^{'2}}$.
Then, if we are interested only in electromagnetic interactions the covariant
derivative becomes:
\bear
D_\mu & = & \partial_\mu - i e
  (T_L^3 +   T_Y)
A_\mu=\partial_\mu + e M_Q A_\mu
\eear
where  $e= g \sin{\theta_W}=g'\cos{\theta_W}$ is the electromagnetic coupling and
$T_L^3+T_Y=i M_Q /2$
is the electromagnetic group $U(1)_{EM}$ generator given by:
\bear
M_Q &=& \left(\begin{array}{ccccc}
0 & -&0&0&0 \\
+ &0&0&0&0 \\
0&0&0 & 0&0 \\
0&0 & 0&0&0 \\
0&0&0&0&0
\end{array}\right).
\nn\\
\eear
Thus the photon field $A_\mu$ only couples to $\omega^1$ and $\omega^2$ or the
complex combination:
\bear
\omega^\pm =\frac{1}{\sqrt{2}}(\omega^1 \mp i \omega^2)
\eear
and by using this covariant derivative the $U(1)_{EM}$ gauge NL$\sigma$M Lagrangian takes the form:
\bear
     \mL^{\rm MCHM}_2(\omega^\alpha,\gamma)
 &=& \Frac{1}{2}\, g_{\alpha\beta}(\omega)\,  D^\mu \omega^\alpha\, D_\mu \omega^\beta
\nn \\
&=& \frac{1}{2}\, g_{\alpha\beta}(\omega)\,  \partial^\mu \omega^\alpha\, \partial_\mu \omega^\beta + ie A_\mu(\omega^-\partial^\mu \omega^+-\omega^+\partial^\mu\omega^-)+   e^2 A^2 \omega^+\omega^-.
\nn\\
\eear
Thus the Higgsless electromagnetic interactions are exactly the same we found in this work
for the ECLh Lagrangian $\mL_2$ in the spherical parametrization of the $S^3$ coset,
\ie   keeping just the constant term $\mF(0)$  in the function $\mF(h)$
and dropping terms with Higgs fields.
   Making  the identification $h=\omega^4$  and $\omega^a$  as in the main text  for  $a=1,2,3$
(see Sec.~\ref{sec.coset}), with $\omega^2= \sum_a (\omega^a)^2$,
%
the Lagrangian becomes
\bear
   \mL^{\rm MCHM}_2(\omega^\alpha,\gamma)
 &=& \Frac{1}{2}\partial^\mu \omega^a\, \partial_\mu \omega^a
   +   \Frac{1}{2}\partial^\mu h \partial_\mu h
      +   \Frac{1}{2} \Frac{ (\omega^+\partial_\mu \omega^-+\omega^-\partial_\mu \omega^+
                      + \omega^0 \partial_\mu \omega^0 + h\partial_\mu h)^2   }{f^2-\omega^2 - h^2}
\nonumber    \\
  & &\qquad\qquad  +  ie A_\mu(\omega^-\partial^\mu \omega^+-\omega^+\partial^\mu\omega^-)
  +   e^2 A^2 \omega^+\omega^-
\nn\\ \nn\\
 &=& \Frac{1}{2}\partial^\mu \omega^a\, \partial_\mu \omega^a
   +   \Frac{1}{2}\partial^\mu h \partial_\mu h
   +   \Frac{1}{2f^2}(\omega^+\partial_\mu \omega^-+\omega^-\partial_\mu \omega^+
                      + \omega^0 \partial_\mu \omega^0 + h\partial_\mu h)^2
\nonumber    \\
  &&  \qquad\qquad  + ie A_\mu(\omega^-\partial^\mu \omega^+-\omega^+\partial^\mu\omega^-)
  +   e^2 A^2 \omega^+\omega^-
  \,\,\, + \,\,\, ...
\nn\\
\eear
 where the dots stand for terms with six or more boson fields, irrelevant for the one-loop
calculation of the photon-photon scattering amplitudes.

\begin{figure}[!t]
\begin{center}
\begin{tabular}{c c c}
\psfig{file=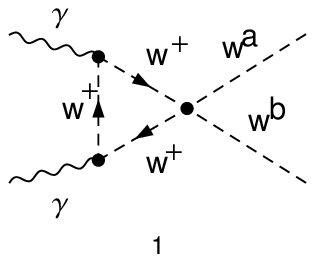,width=2.5cm,clip=} & \psfig{file=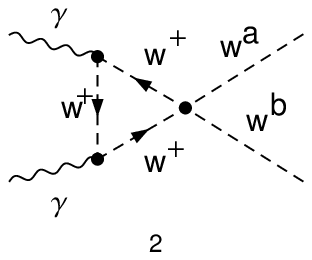,width=2.5cm,clip=}
& \psfig{file=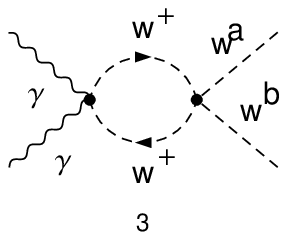,width=2.5cm,clip=}
\end{tabular}
\caption{{\small  MCHM one-loop diagrams for $\gamma\gamma\to w^a w^b$. }}
\label{fig.MCHM}
\end{center}
\end{figure}

Now the point is that for the computation of the one-loop
$\gamma\gamma \to zz$ and $\gamma\gamma\to w^+ w^-$  amplitudes we can use
the $S^4$ gauged NL$\sigma$M Lagrangian above which has a much simpler structure than the ones used in the main text.
Then according to  \cite{DobadoMorales},
where these processes were considered in the framework of general
$SO(N+1)/SO(N)$ gauged NL$\sigma$M
 for low-energy QCD,
the one-loop computation only involves the  bubble and triangle diagrams
(Fig.~\ref{fig.MCHM})   which are very easy to compute. The result is simply
\bear
A(s,t,u)^{\gamma\gamma\to zz} &=& \,-\, \Frac{1}{4\pi^2 f^2}\,=  \,-\, \Frac{(1-a^2)}{4\pi^2 v^2}\, ,
\nn\\
A(s,t,u)^{\gamma\gamma\to w^+ w^-} &=& \,-\, \Frac{1}{8\pi^2 f^2}\, = \,-\,\Frac{(1-a^2)}{8\pi^2 v^2},
\nn\\
\label{eq.results2}
\eear
   where we have used the relation $(1-a^2)=v^2/f^2$ between $f$, $v$ and $a$
from $SO(5)/SO(4)$ MCHM~\cite{MCHM}.

 If instead of using the $\Phi$ vector representation in Eq.~\eqref{eq.Phi-vector}
for the $SO(5)/SO(4)$ Goldstone bosons  we employ the exponential parametrization
in ref.~\cite{MCHM} it is not difficult to check that $\mL_2^{\rm MHCM}$
has the same structure and  produces the same one-loop photon-photon amplitudes
as  the general ECLh Lagrangian $\mL_2$ considered in the main text provided
\bear
a^2 \,=\, \cos^2\theta \, =\,  1-\Frac{v^2}{f^2}\, ,\qquad\qquad \qquad
b\, =\, \cos(2\theta) \,= \, 1\, -\, 2\, \Frac{v^2}{f^2}\, .
\eear
That means that if we are interested only in processes which do not depend
on the $b$ parameter appearing in ${\cal F}(h)$,
the results obtained from the $SO(5)/SO(4)$ MCHM are universal
(we only need to tune $f$ and $\sin{\theta}$ to get the required $a$ parameter
according to the previous equation). This is in particular the case
of the processes considered in this work,  $\gamma\gamma\to w^+ w^-$ and $\gamma\gamma\to zz$.
However this is not the case for other kind of processes as for example
$w^aw^b \to w^c w^d$, $w^a w^b \to hh $ or $hh \to hh$ considered in \cite{Dobado:2013}.

\section{Related observables: $S$--parameter and other photon  transitions}
\label{app.related-obs}

In this work we computed the $\gamma\gamma\to  zz$ and $\gamma\gamma\to  w^+w^-$
scattering amplitudes up to NLO in the chiral expansion. It is not difficult
to find other simple observables
where the bosonic contribution is determined
 by the same effective parameters.
 We remind the reader that the
fermionic contributions
  is not considered here. It must be eventually taken into account in    a realistic
 phenomenological analysis of the experimental data.
In this appendix we discuss six of these observables described in terms of four
independent combinations  of couplings,
$a$, $a_1^r$, $(a_2^r-a_3^r)$, $c_\gamma^r$ (see Table~\ref{tab.relevant-couplings}
for a summary):
\begin{itemize}

\item{\bf $\gamma\gamma\to zz$ and $\gamma\gamma\to w^+w^-$ scattering amplitudes:}
\\ \\
 These are the main results
of this work  (Eqs.~\eqref{eq.A-zz}--\eqref{eq.B-zz} and~\eqref{eq.A-ww}--\eqref{eq.B-ww}).
The total one-loop amplitude has been found here to be UV finite and the relevant $\cO(p^4)$
  ECLh parameters
involved in these processes are found to be renormalization group invariant.

\item{\bf  $\Gamma(h\to \gamma\gamma)$:}
\\ \\
  We have computed the  one-loop       bosonic contribution to the
$h(q)\to \gamma(k_1,\epsilon_1)\gamma(k_2,\epsilon_2)$   decay width from the ECLh.
Under the same approximations   considered in the text this is given by the amplitude
(fermion loops are absent),
\bear
\mM_{h\to \gamma\gamma} &=&
  \Frac{ie^2}{v}  \,\,\,  \left(\, m_h^2\, (\epsilon_1\epsilon_2)\, -\, 2 \, (k_2\epsilon_1)\,  (k_1\epsilon_2)\,\right)
\,\,\,   \left[ \, c_\gamma^r \, +\, \Frac{a}{8\pi^2}\,\right]  \, .
\label{eq.M-h-gamgam}
\eear
If the fermionic contributions are dropped one has the      
  following modification with respect to the SM result,
\bear
\Gamma(h\to\gamma\gamma)  &=&
\Gamma(h\to\gamma\gamma)^{\rm SM}
   \,\,\, \left[\, a\, + \, 8\pi^2 c_\gamma^{r}\, \right]^2 \, ,
\label{eq.width-h-gamgam}
\eear
with  $\Gamma(h\to\gamma\gamma)^{\rm SM} =\frac{\alpha^2 m_h^3}{64\pi^3 v^2}$.
Higher order terms in the $m_h^2/(16\pi^2 v^2)$ expansion
have been dropped in the latter and previous two equations.
Likewise,   we are just keeping the  lowest order in the $g^{(')}$ expansion
($\cO(\alpha)$ in Eq.~\eqref{eq.M-h-gamgam} and $\cO(\alpha^2)$ in Eq.~\eqref{eq.width-h-gamgam})
and not including higher order  corrections (for $v$ fixed).
The total one-loop amplitude is found again to be UV--finite and hence no renormalization
is needed for the $\cO(p^4)$     ECLh parameter    $c_\gamma$.
It is then trivial  to check that for $a=1$ and $c_\gamma^r=0$  one recovers the SM result
for $\alpha^{-2}\Gamma(h\to\gamma\gamma)$  in the limit $g,g'\to 0$ and without the  fermion
loop contributions~\cite{h-gam+gam}.


\item{\bf Oblique $S$--parameter}
\\ \\
The first non-vanishing contribution appears at   NLO.     
Likewise, we find that the one-loop amplitude is  UV--divergent   
and needs to be renormalized by means of the
ECLh parameter $a_1$. In the $\overline{MS}$ scheme we find
\bear
S &=&
\, -\, 16 \pi a_1^r
\,\,\, +\,\,\, \Frac{(1-a^2)}{12 \pi}\, \bigg( \,\Frac{5}{6}\, + \, \ln\Frac{\mu^2}{m_h^2}\,\bigg)\, ,
\label{eq.S-prediction}
\eear
with $a_1^r$ being the renormalized coupling at the  renormalization scale $\mu$,
and absorbing the UV--divergences from the one-loop diagrams.
In this expression,  the oblique parameter is defined with
the reference value $m_h^{\rm Ref}$
set to the physical Higgs mass~\cite{Peskin:92}.
   Notice that the NLO from the Higgsless ECL~\cite{Dobado:99} is again recovered for $a=0$.
Likewise, we recover the $(1-a^2)$ coefficient of the
logarithm from the one-loop computation~\cite{Pich:2013} with a Chiral Lagrangian
including also vector and axial-vector resonances.

\item{\bf Electromagnetic vector form-factor ($\gamma^*\to w^+w^-$)}
\\ \\
The electromagnetic transition $\gamma^*\to w^+w^-$ from a virtual
photon with momentum $q^\mu=p_1^\mu+p_2^\mu$  is described through the matrix element
\bear
\bra w^+(p_1) \,w^-(p_2)|\,  J^\mu_{\rm EM} \, |0\ket &=&
 \,   e  \,
 (p_1^\mu-p_2^\mu)\, \mF_{\gamma^*ww}(q^2)\, .
\eear
\ The electromagnetic vector form-factor (VFF) can be computed  with  the ECLh
up to NLO. We find
\bear
\mF_{\gamma^*ww} &=& 1
 +\Frac{ 2 q^2       ( a_3^r-a_2^r)         }{v^2}
+    (1-a^2)\Frac{q^2 }{96\pi^2 v^2}\left(
\Frac{8}{3}   -\ln\Frac{-q^2}{\mu^2}\right),
\eear
with the $\cO(p^4)$ chiral couplings given   in the $\overline{MS}$ scheme
at the scale $\mu$ and renormalizing the one-loop UV--divergences.


\item{\bf Higgs transition form-factor ($\gamma^*\gamma^* \to h$)}
\\ \\
An interesting observable in order to pin down the $h\gamma\gamma$ coupling $c_\gamma^r$
is the  Higgs transition form-factor
  (HTFF),
which describes the process $\gamma^*(k_1)\gamma^*(k_2)\to h(p)
$~\cite{Higgs-TFF,Higgs-TFF-Sanz,Higgs-TFF-Trott}.
This transition is given by the matrix element,
\bear
\Int d^4x\, e^{-i k_1x} \, \bra h(p)|\, T\{ \, J_{\rm EM}^\mu(x)\,  J_{\rm EM}^\nu(0)\, \}
\, |0\ket
&=& \,i
  \, e^2  \,
\,\left[\, (k_1k_2) \, g^{\mu\nu}\,- \, k_2^\mu \, k_1^\nu  \, \right]
\,\,\mF_{\gamma^*\gamma^* h}(k_1^2,k_2^2)\, ,
\nn\\
\eear
where in the case when one of the photons is on-shell
   we find that the $\cO(p^4)$ HTFF is given by
\bear
\mF_{\gamma^*\gamma h}(k^2, 0)  &=&  \, -\, 2 \,
  c_\gamma^{r} \, ,
\eear
with no contribution present at $\cO(p^2)$
   nor coming from loops at $\cO(p^4)$.
Here again we provide the result in the $m_h\to 0$ limit
and higher correction of $\cO(m_h^2/(16\pi^2 v^2))$
have been dropped.
Notice that this result does not correspond to
the same kinematical regime as $\Gamma(h\to\gamma\gamma)$,
since here we are considering  $m_h^2\ll k^2\ll 16\pi^2 v^2$. The one-loop
 $\cO(p^4)$
diagrams cancel out completely in this energy range and, therefore,
there are no UV--divergences and again  $c_\gamma$ does not need to be renormalized.

\end{itemize}

\end{document}